\documentclass[useAMS]{mn2e}
\usepackage{afterpage}
\usepackage{float}
\usepackage{longtable}
\usepackage{times}
\usepackage[]{natbib}
\usepackage{multirow}
\usepackage[dvips]{graphicx}
\usepackage{rotating}

\def\aj{AJ}                   
\def\apj{ApJ}                 
\def\apjl{ApJ}                
\def\aap{A\&A}                
\def\mnras{MNRAS}             
\def\nat{Nature}              
\def\physrep{Phys.~Rep.}      

\font \bolditalics = cmmib10
\def \vc #1{{\textfont1=\bolditalics \hbox{$\bf#1$}}}
\def\pct{\%}
\def\rg{{\bf r}}
\def\gammag{{\vc \gamma}}
\def\Cg{{\bf C}}
\def\mg{{\bf m}}
\def\xig{{\vc \xi}}

\begin{document}
\title[100 deg$^2$ weak lensing survey]{Cosmological Constraints From the 100 Square Degree Weak Lensing Survey\thanks{Based on observations obtained at the Canada-France-Hawaii Telescope (CFHT) which is operated by the National Research Council of Canada (NRCC), the Institut des Sciences de l'Univers (INSU) of the Centre National de la Recherche Scientifique (CNRS) and
the University of Hawaii (UH)}}

\author[Benjamin et al.]{Jonathan Benjamin$^{1}$\thanks{jonben@phas.ubc.ca}, Catherine Heymans$^{1,2}$, Elisabetta Semboloni$^{2,3}$,\newauthor Ludovic Van Waerbeke$^{1}$, Henk Hoekstra$^{4}$, Thomas Erben$^{3}$, Michael D. Gladders$^{5}$, \newauthor Marco Hetterscheidt$^{3}$, Yannick Mellier$^{2,3,6}$ \& H.K.C. Yee$^{7}$\\
	$^1$ University of British Columbia, 6224 Agricultural Road, Vancouver, V6T 1Z1, B.C., Canada.\\
	$^2$ Institut d'Astrophysique de Paris, UMR7095 CNRS, Universit\'e Pierre~\&~Marie Curie - Paris, 98 bis bd Arago, 75014 Paris, France.\\
	$^3$ Argelander-Institut f\"{u}r Astronomie (AIfA), Universit\"{a}t Bonn, Auf dem H\"{u}gel 71, 53121 Bonn, Germany.\\	
	$^4$ Department of Physics and Astronomy, University of Victoria, Victoria, BC, V8P 5C2, Canada.\\
	$^5$ Department of Astronomy \& Astrophysics, University of Chicago, Chicago, IL 60637, USA \\
	$^6$ Observatoire de Paris, LERMA, 61, avenue de l'Observatoire, 75014 Paris, France.\\
	$^7$ Department of Astronomy \& Astrophysics, University of Toronto, Toronto, ON, M5S 3H4, Canada}

\date{22 August 2007}

\maketitle

\begin{abstract}
We present a cosmic shear analysis of the 100 square degree weak lensing survey, combining data from the CFHTLS-Wide, RCS, VIRMOS-DESCART and GaBoDS surveys.  Spanning $\sim100$ square degrees, with a median source redshift $z \sim 0.78$, this combined survey allows us to place tight joint constraints on the matter density parameter $\Omega_\mathrm{m}$, and the amplitude of the matter power spectrum $\sigma_\mathrm{8}$, finding $\sigma_\mathrm{8} (\frac{\Omega_\mathrm{m}}{0.24})^{0.59} = 0.84\pm0.05$.  Tables of the measured shear correlation function and the calculated covariance matrix for each survey are included as Supplementary Material to the online version of this article. 

The accuracy of our results are a marked improvement on previous work owing to three important differences in our analysis; we correctly account for sample variance errors by including a non-Gaussian contribution estimated from numerical simulations;  we correct the measured shear for a calibration bias as estimated from simulated data; we model the redshift distribution, $n(z)$, of each survey from the largest deep photometric redshift catalogue currently available from the CFHTLS-Deep.   This catalogue is randomly sampled to reproduce the magnitude distribution of each survey with the resulting survey dependent $n(z)$ parametrised using two different models.  While our results are consistent for the $n(z)$ models tested, we find that our cosmological parameter constraints depend weakly (at the $5\pct$ level) on the inclusion or exclusion of galaxies with low confidence photometric redshift estimates ($z>1.5$).   These high redshift galaxies are relatively few in number but contribute a significant weak lensing signal.   It will therefore be important for future weak lensing surveys to obtain near-infra-red data to reliably determine the number of high redshift galaxies in cosmic shear analyses. 
\end{abstract}

\begin{keywords}
 cosmology: cosmological parameters - gravitational lenses - large-scale structure of Universe - observations
\end{keywords}

\section{Introduction}
Weak gravitational lensing of distant galaxies by intervening matter provides a unique and unbiased tool to study the matter distribution of the Universe.  Large scale structure in the Universe induces a weak lensing signal, known as cosmic shear.  This signal provides a direct measure of the projected matter power spectrum over a redshift range determined by the lensed sources and over scales ranging from the linear to non-linear regime. The unambiguous interpretation of the weak lensing signal makes it a powerful tool for measuring cosmological parameters which complement those from other probes such as, CMB anisotropies \citep{spergel-2006} and type Ia supernovae \citep{2006AA...447...31A}.

Cosmic shear has only recently become a viable tool for observational cosmology, with the first measurements reported simultaneously in 2000 \citep{2000MNRAS.318..625B,2000astro.ph..3338K,2000A&A...358...30V,2000Natur.405..143W}. The data these early studies utilised were not optimally suited for the extraction of a weak lensing signal. Often having a poor trade off between sky coverage and depth, and lacking photometry in more than one colour, the ability of these early surveys to constrain cosmology via weak lensing was limited, but impressive based on the data at hand.  Since these early results several large dedicated surveys have detected weak lensing by large-scale structure, placing competitive constraints on cosmological parameters   \citep{2002ApJ...577..595H,2003MNRAS.344..673B,2003AJ....125.1014J,2003MNRAS.341..100B,2003ApJ...597...98H,2005MNRAS.359.1277M,2004ApJ...605...29R,2005AA...429...75V,2005MNRAS.361..160H,2006AA...452...51S,2006ApJ...647..116H,2006astro.ph..6611S}. With the recent first measurements of a changing lensing signal as a function of redshift \citep{2005MNRAS.363..723B,2005ApJ...632L...5W,2006AA...452...51S,2007astro.ph..1480M}, and the first weak lensing constraints on dark energy \citep{2006ApJ...644...71J,2006astro.ph.10284K,2006ApJ...647..116H,2007A&A...463..405S}, the future of lensing is promising.

There are currently several `next generation' surveys being conducted. Their goal is to provide multi-colour data and excellent image quality, over a wide field of view.  Such data will enable weak lensing to place tighter constraints on cosmology, breaking the degeneracy between the matter density parameter $\Omega_\mathrm{m}$ and the amplitude of the matter power spectrum $\sigma_\mathrm{8}$ and allowing for competitive constraints on dark energy.  In this paper, we present an analysis of the 100 deg$^2$ weak lensing survey that combines data from four of the largest surveys analysed to date, including the wide component of the Canada-France-Hawaii Telescope Legacy Survey \citep[CFHTLS-Wide,][]{2006ApJ...647..116H}, the Garching-Bonn Deep Survey \citep[GaBoDs,][]{Hetterscheidt..0606571}, the Red-Sequence Cluster Survey \citep[RCS,][]{2002ApJ...577..595H}, and the VIRMOS-DESCART survey \citep[VIRMOS,][]{2005AA...429...75V}.  These surveys have a combined sky coverage of 113 deg$^2$ (96.5 deg$^2$, after masking), making this study the largest of its kind.

Our goal in this paper is to provide the best estimates of cosmology currently attainable by weak lensing, through a homogeneous analysis of the major data sets available. Our analysis is distinguished from previous work in three important ways. For the first time we account for the effects of non-Gaussian contributions to the analytic estimate of the sample variance -- sometimes referred to as cosmic variance -- covariance matrix \citep{2002A&A...396....1S}, as described in \citet{Semboloni..0606648}. Results from the Shear TEsting Programme \citep[STEP, see][]{2007MNRAS.376...13M, 2006MNRAS.368.1323H} are used to correct for calibration error in our shear measurement methods, a marginalisation over the uncertainty in this correction is performed. We use the largest deep photometric redshift catalogue currently available \citep{Ilbert..0603217}, which provides redshifts for $\sim500000$ galaxies in the CFHTLS-Deep.  Additionally, we properly account for sample variance in our calculation of the redshift distribution, an important source of error discussed in \citet{Waerbeke..0603696}.

This paper is organised as follows: in \S2 we give a short overview of cosmic shear theory, and outline the relevant statistics used in this work. We describe briefly each of the surveys used in this study in \S3. We present the measured shear signal in \S4. In \S5 we present the derived redshift distributions for each survey. We present the results of the combined parameter estimation in \S6, closing thoughts and future prospects are discussed in \S7.

\section{Cosmic Shear Theory}
\label{sec:theory}

We briefly describe here the notations and statistics used in our cosmic shear analysis. For detailed
reviews of weak lensing theory the reader is referred to \citet{2006astro.ph.12667M},
\citet{2001PhR...340..291B}, and \citet{1998MNRAS.296..873S}. The notations used in the
latter are adopted here.
The power spectrum of the projected density field (convergence $\kappa$) can be written as
\begin{eqnarray}
P_\kappa(k)&=&\frac{9\,\rm{H}_0^4\,\Omega_{\mathrm{m}}^2}{4c^4}\int_0^{w_H} {{\rm d}w \over a^2(w)}
P_{3D}\left({k\over f_K(w)};
w\right)\times\nonumber\\
&&\left[ \int_w^{w_H}{\rm d} w' n(w') {f_K(w'-w)\over f_K(w')}\right]^2,
\label{pofkappa}
\end{eqnarray}
\noindent where $\rm{H}_0$ is the Hubble constant, 
$f_K(w)$ is the comoving angular diameter distance
out to a distance $w$ ($w_H$ is the comoving horizon distance), 
$a(w)$ is the scale factor, and
$n[w(z)]$ is the redshift distribution of the sources (see \S5).
$P_{3D}$ is the 3-dimensional mass power spectrum computed from a
non-linear estimation of the dark matter clustering
\citep[see for example][]{1996MNRAS.280L..19P,2003MNRAS.341.1311S}, 
and $k$ is the 2-dimensional wave vector perpendicular to the line-of-sight. 
$P_{3D}$ evolves with time, hence it's dependence on the co-moving radial coordinate $w$.

In this paper we focus on the shear correlation function statistic $\xi$.  
For a galaxy pair separation $\theta$, we define
\begin{eqnarray}
\xi_+(\theta)&=&
\langle \gamma_t(r)\gamma_t(r+\theta)\rangle+\langle \gamma_r(r)\gamma_r(r+\theta)
\rangle.\nonumber \\
\xi_-(\theta)&=&\langle \gamma_t(r)\gamma_t(r+\theta)\rangle -\langle \gamma_r(r)\gamma_r(r+\theta)\rangle,
\label{xipm}
\end{eqnarray}
where the shear $\gammag = (\gamma_t, \gamma_r)$ is rotated into 
the local frame of the line joining the centres of each galaxy pair 
separated by $\theta$. The tangential shear is $\gamma_t$, while $\gamma_r$
is the rotated shear (having an angle of $\pi \over 4$ to the tangential component).
The shear correlation function $\xi_+$ is related to the convergence power 
spectrum through
\begin{equation}
\xi_+(\theta)={1\over 2\pi} \int_0^\infty~{\rm d} k~
 k P_\kappa(k) J_0(k\theta),
\label{theogg}
\end{equation}
\noindent where $J_0$ is the zeroth order Bessel function of the first kind.
A quantitative measurement of the lensing amplitude and the
systematics is obtained by splitting the signal into its
curl-free ($E$-mode) and curl ($B$-mode) components respectively.
This method has been advocated to help the measurement of the
intrinsic alignment contamination in the weak lensing signal
\citep{2001ApJ...559..552C,2002ApJ...568...20C}, but it is also an
efficient measure of the residual systematics from the PSF
correction \citep{2002ApJ...567...31P}.

The $E$ and $B$ modes derived from the shape of galaxies are
unambiguously defined only for the so-called aperture mass variance
$\langle M_{\rm ap}^2\rangle$, which is a weighted shear variance
within a cell of radius $\theta_c$. The cell itself is defined as a
compensated filter \citep{1998MNRAS.296..873S}, such that a constant convergence $\kappa$ gives
$M_{\rm ap}=0$. $\langle M_{\rm ap}^2\rangle$ can be rewritten as a
function of the tangential shear $\gamma_t$ if we express
$\gammag=(\gamma_t,\gamma_r)$ in the local frame of the line
connecting the aperture centre to the galaxy. $\langle M_{\rm ap}^2 \rangle$ is given by:
\begin{equation}
\langle M_{\rm ap}^2\rangle={288\over \pi\theta_c^4} \int_0^\infty~{{\rm d}k\over k^3}
 P_\kappa(k) [J_4(k\theta_c)]^2.
\label{theomap}
\end{equation}
The $B$-mode $\langle M_{\rm ap}^2\rangle_\perp$ is obtained by
replacing $\gamma_t$ with $\gamma_r$. The aperture mass is
insensitive to the mass sheet degeneracy, and therefore it provides
an unambiguous splitting of the $E$ and $B$ modes. The drawback is
that aperture mass is a much better estimate of the small scale
power than the large scale power which can be seen from the function
$J_4(k\theta_c)$ in Eq.(\ref{theomap}) which peaks at $k\theta_c\sim
5$. Essentially, all scales larger than a fifth of the largest
survey scale remain inaccessible to $M_{\rm ap}$. The large-scale
part of the lensing signal is lost by $M_{\rm ap}$, while the
remaining small-scale fraction is difficult to interpret because the
strongly non-linear power is difficult to predict accurately
\citep{2002A&A...393..369V}.  
It is therefore preferable to decompose the shear correlation function 
into its $E$ and $B$ modes, 
as it is a much deeper probe of the linear regime.  

Following \citet{2001ApJ...559..552C, 2002ApJ...568...20C}, 
we define

\begin{equation}
\xi'(\theta)=\xi_-(\theta)+4\int_\theta^\infty \frac{{\rm d}\vartheta}{\vartheta} \xi_-(\vartheta)
	-12\theta^2 \int_\theta^\infty \frac{{\rm d}\vartheta}{\vartheta^3}\xi_-(\vartheta).
\label{eqn:xipr}
\end{equation}
The $E$ and $B$ shear correlation functions are then given by
\begin{equation}
\xi_E(\theta)=\frac{\xi_+(\theta)+\xi'(\theta)}{2},\ \ \ \ \ \
\xi_B(\theta)=\frac{\xi_+(\theta)-\xi'(\theta)}{2}.
\label{eqn:xieb}
\end{equation}
In the absence of systematics, $\xi_B = 0$ and $\xi_E = \xi_+$ 
(Eq.\ref{theogg}). 
In contrast with the $M_{\rm ap}$ statistics, the separation of the
two modes depends on the signal integrated 
out with the scales probed by all surveys.  One option is to calculate 
Eq.(\ref{eqn:xieb}) using a fiducial cosmology to compute $\xi_-$ on scales $\theta \rightarrow \infty$.  
As shown in \citet{2005MNRAS.361..160H}, changes in the choice of fiducial cosmology 
do not significantly affect the results, allowing this statistic 
to be used as a diagnostic tool for the presence of systematics.  This option 
does however prevent us from using $\xi_E$ to constrain cosmology.  
As the statistical noise on the measured $\xi_E$ is $\sim \sqrt{2}$ smaller 
than the statistical noise on the measured $\xi_+$, there is a more preferable alternative.
If the survey size is sufficiently large in comparison to the scales probed 
by $\xi_E$, we can consider the unknown integral to be a constant that 
can be calibrated using the aperture mass $B$-mode statistic $\langle M_{\rm ap}^2\left(\Delta\theta\right)\rangle_\perp$.  
A range $\Delta\theta$ of angular scales where $\langle M_{\rm ap}^2\left(\Delta\theta\right)\rangle_\perp\sim 0$ ensures that the $B$-mode of the shear correlation function is zero as well (within the error bars), at angular scales $\sim \frac{\Delta\theta}{5}$.   In this analysis we have calibrated $\xi_{E,B}$ for our survey data in this manner.  This alternative method has also been verified by re-calculating $\xi_{E,B}$ using our final best-fit cosmology to extrapolate the signal.  We find the two methods to be in very close agreement.

\section{Data description}
\begin{table*}
\caption{Summary of the published results from each survey used in this study. The values for $\sigma_\mathrm{8}$ correspond to $\Omega_\mathrm{m} = 0.24$, and are given for the \protect\citet{1996MNRAS.280L..19P} method of calculating the non-linear evolution of the matter power spectrum. The statistics listed were used in the previous analyses; the shear correlation functions $\xi_{E,B}$ (Eq.~\ref{eqn:xieb}), the aperture mass statistic $\langle M_{\rm ap}^2 \rangle$ (Eq.~\ref{theomap}) and the top-hat shear variance $\langle \gamma^2_{E,B}\rangle$ (see \protect\citet{2005AA...429...75V}).}
\label{tab:summary}
\begin{center}
\begin{tabular}{lllll}
\hline
\hline
                     & CFHTLS-Wide & GaBoDS & RCS & VIRMOS-DESCART\\
\hline
Area ($\rm{deg}^2$)  & 22.0 & 13.0   & 53  & 8.5\\
N$_{\rm fields}$     & 2 & 52     & 13  & 4\\
Magnitude Range      & $21.5 < i' < 24.5$ & $21.5 < R < 24.5$ & $22 < R_\mathrm{C} < 24$ & $21 < I_\mathrm{AB} < 24.5$\\
$<z_{\rm source}>$   & $0.81$ & $0.78$ & $0.6$ & $0.92$\\
$z_{\rm median}$           & $0.71$ & $0.58$ & $0.58$ & $0.84$\\
Previous Analysis    & \citet{2006ApJ...647..116H} & \citet{Hetterscheidt..0606571} & \citet{2002ApJ...577..595H} & \citet{2005AA...429...75V}\\
$\sigma_\mathrm{8}$ ($\Omega_\mathrm{m}=0.24$)  & $0.99\pm 0.07$ & $0.92\pm 0.13$ & $0.98\pm 0.16$ & $0.96\pm 0.08$\\
Statistic            & $\xi_{E,B}(\theta)$ & $\langle M_{ap}^2 \rangle(\theta)$ & $\langle M_{ap}^2 \rangle(\theta)$ & $\langle \gamma^2_{E,B} \rangle(\theta)$\\
\hline
\hline
\end{tabular}
\end{center}
\end{table*}
In this section we summarise the four weak lensing surveys that form the 100 deg$^2$ weak lensing survey: CFHTLS-Wide \citep{2006ApJ...647..116H}, GaBoDs \citep{Hetterscheidt..0606571}, RCS \citep{2002ApJ...577..595H}, and VIRMOS-DESCART \citep{2005AA...429...75V}, see Table~\ref{tab:summary} for an overview. Note that we have chosen not to include the CFHTLS-Deep data \citep{2006AA...452...51S} in this analysis  since the effective area of the analysed data is only 2.3 deg$^2$. Given the breadth of the parameter constraints obtained from this preliminary data set there is little value to be gained by combining it with four surveys which are among the largest available.

\subsection{CFHTLS-Wide}
The Canada-France-Hawaii Telescope Legacy Survey (CFHTLS) is a joint Canadian-French program designed to take advantage of Megaprime, the CFHT wide-field imager.  This 36 CCD mosaic camera has a $1 \times 1$ degree field of view and a pixel scale of 0.187 arcseconds per pixel.  The CFHTLS has been allocated $\sim450$ nights over a 5 year period (starting in 2003), the goal is to complete three independent survey components; deep, wide, and very wide. The wide component will be imaged with five broad-band filters: \textit{u}$^*$, \textit{g}', \textit{r}', \textit{i}', \textit{z}'. With the exception of \textit{u}$^*$ these filters are designed to match the SDSS photometric system \citep{1996AJ....111.1748F}.
The three wide fields will span a total of 170 deg$^2$ on completion. In this study, as in \citet{2006ApJ...647..116H}, we use data from the W1 and W3 fields that total 22 deg$^2$ after masking, and reach a depth of $i'=24.5$.  For detailed information concerning data reduction and object analysis the reader is referred to \citet{2006ApJ...647..116H}, and \citet{2002A&A...393..369V}.

\subsection{GaBoDS}
The Garching-Bonn Deep Survey (GaBoDS) data set was obtained with the Wide Field Imager (WFI) of the MPG/ESO 2.2m telescope on La Silla, Chile. The camera consists of 8 CCDs with a 34 by 33 arcminute field of view and a pixel scale of 0.238 arcseconds per pixel.   The total data set consists of 52 statistically independent fields, with a total effective area (after trimming) of $13\,{\rm deg}^2$. The magnitude limits of each field ranges between $R_{\rm VEGA} = 25.0$ and $R_{\rm VEGA} = 26.5$ (this corresponds to a $5\sigma$ detection in a 2 arcsecond aperture).  To obtain a homogeneous depth across the survey we follow \citet{Hetterscheidt..0606571} performing our lensing analysis with only those objects which lie in the complete magnitude interval $R \in [21.5,24.5]$.  For further details of the data set, and technical information regarding the reduction pipeline the reader is referred to \citet{Hetterscheidt..0606571}.

\subsection{VIRMOS-DESCART}
The DESCART weak lensing project is a theoretical and observational program for cosmological weak lensing investigations. The cosmic shear survey carried out by the DESCART team uses the CFH12k data jointly with the VIRMOS survey to produce a large homogeneous photometric sample in the \textit{BVRI} broad band filters. CFH12k is a 12 CCD mosaic camera mounted at the Canada-France-Hawaii Telescope prime focus, with a 42 by 28 arcminute field of view and a pixel scale of 0.206 arcseconds per pixel. 
The VIRMOS-DESCART data consist of four uncorrelated patches of 2 x 2 deg$^2$ separated by more than 40 degrees. 
Here, as in \citet{2005AA...429...75V}, we use $I_\mathrm{AB}$-band data from all four fields, which have an effective area of 8.5 deg$^2$ after masking, and a limiting magnitude of $I_{\mathrm{AB}}=24.5$ (this corresponds to a $5\sigma$ detection in a 3 arcsecond aperture). 
Technical details of the data set are given in \citet{2001A&A...374..757V} and \citet{2003A&A...410...17M}, while an overview of the VIRMOS-DESCART survey is given in \citet{2004A&A...417..839L}.

\subsection{RCS}
The Red-Sequence Cluster Survey (RCS) is a galaxy cluster survey designed to provide a large sample of optically selected clusters of galaxies with redshifts between 0.3 and 1.4. The final survey covers 90 deg$^2$ in both $R_\mathrm{C}$ and $z'$, spread over 22 widely separated patches of $\sim 2.1\times 2.3$ degrees. The northern half of the survey was observed using the CFH12k camera on the CFHT, and the southern half using the Mosaic II camera mounted at the Cerro Tololo Inter-American Observatory (CTIO), Victor M. Blanco 4m telescope prime focus. This camera has an 8 CCD wide field imager with a 36 by 36 arcminute field of view and a pixel scale of 0.27 arcseconds per pixel.
The data studied here, as in \citet{2002ApJ...577..595H}, consist of a total effective area of 53 deg$^2$ of masked imaging data, spread over 13 patches, with a limiting magnitude of 25.2 (corresponding to a $5\sigma$ point source depth) in the $R_\mathrm{C}$ band. 43 deg$^2$ were taken with the CFHT, the remaining 10 deg$^2$ were taken with the Blanco telescope. 
A detailed description of the data reduction and object analysis is described in \citet{2002ApJ...572...55H}, to which we refer for technical details. 

\section{Cosmic shear signal}
\begin{figure*}
\begin{center}
\vbox{
\hbox{
\includegraphics[scale=0.33]{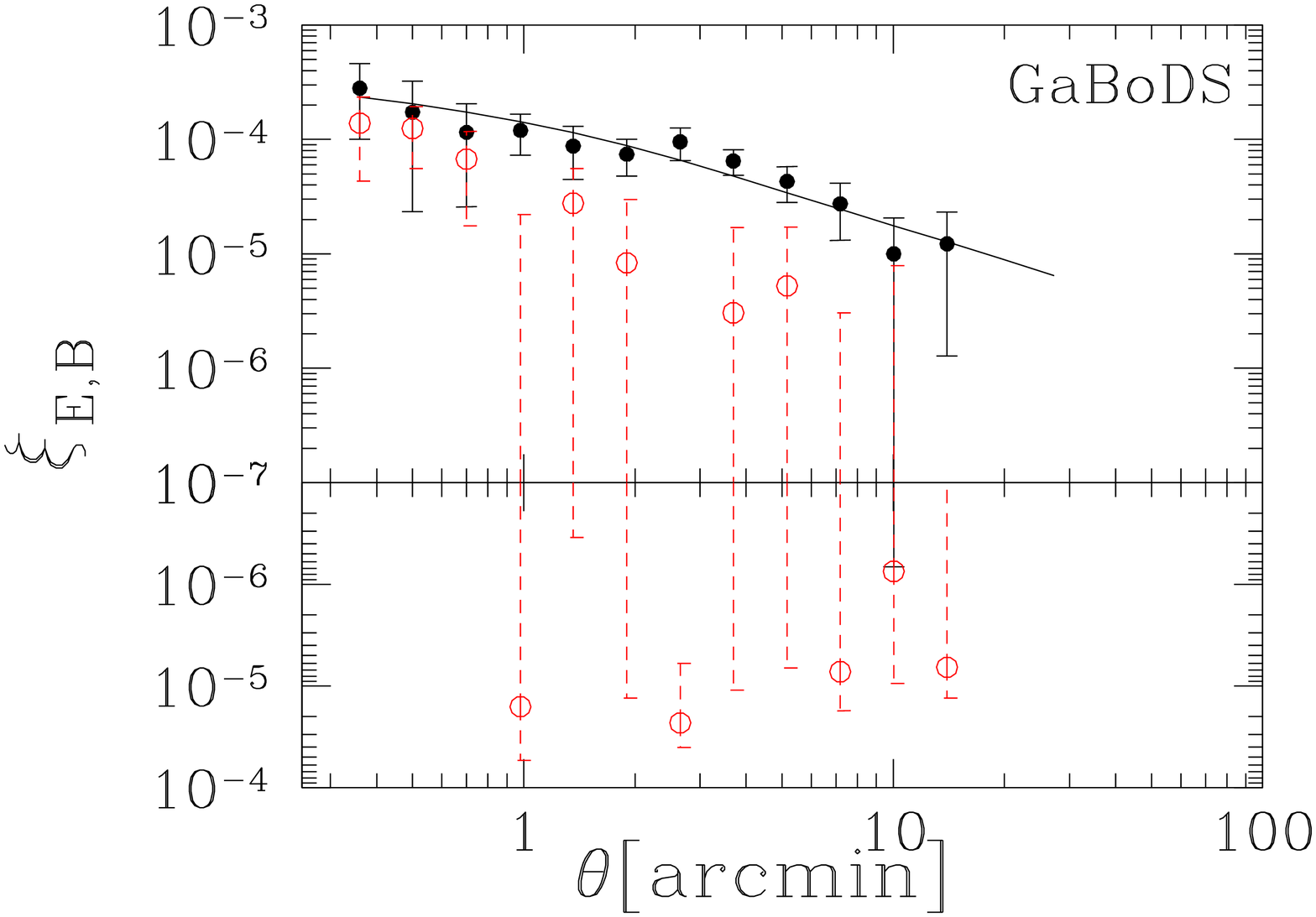}
\includegraphics[scale=0.33]{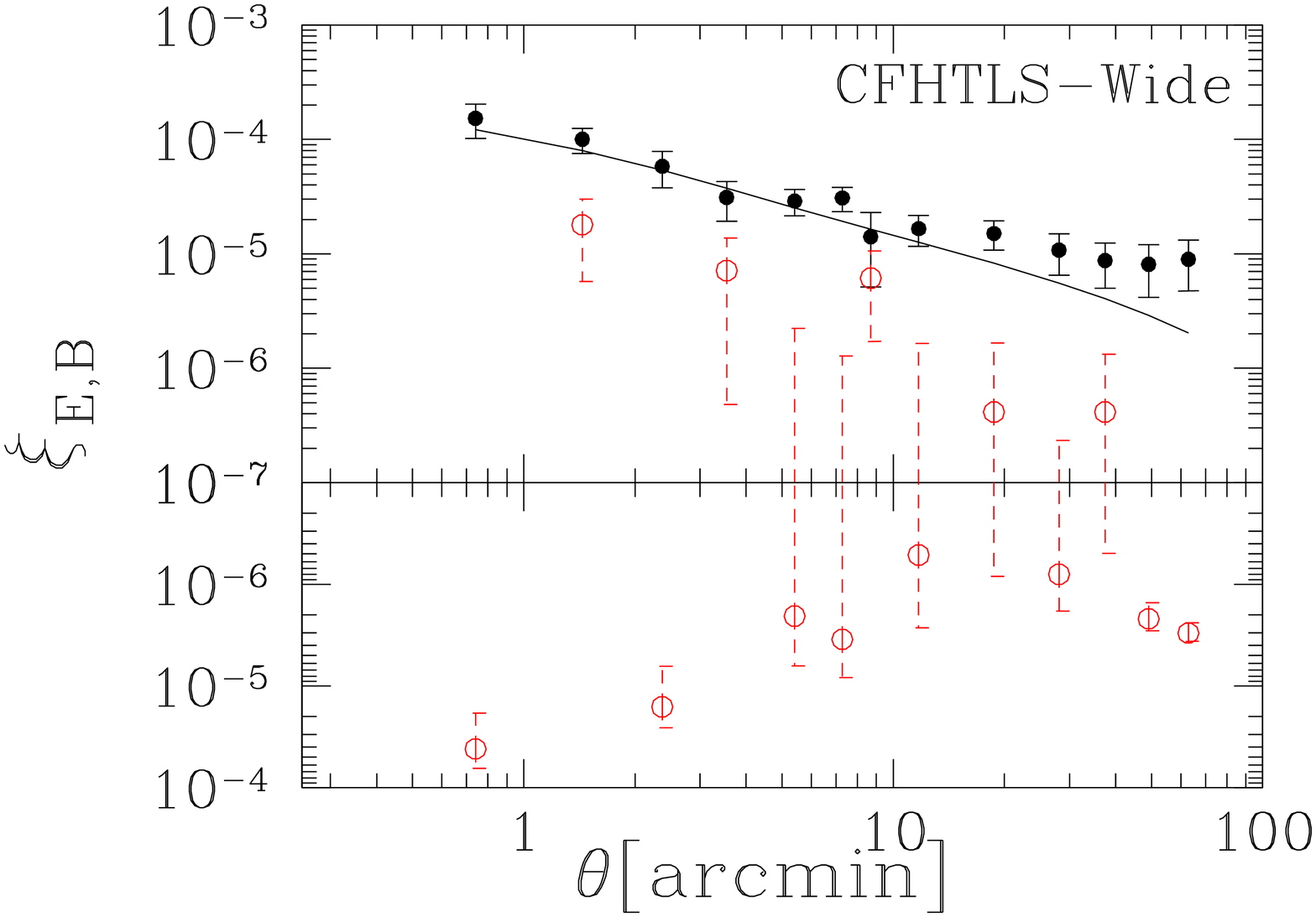}
}
\hbox{
\includegraphics[scale=0.33]{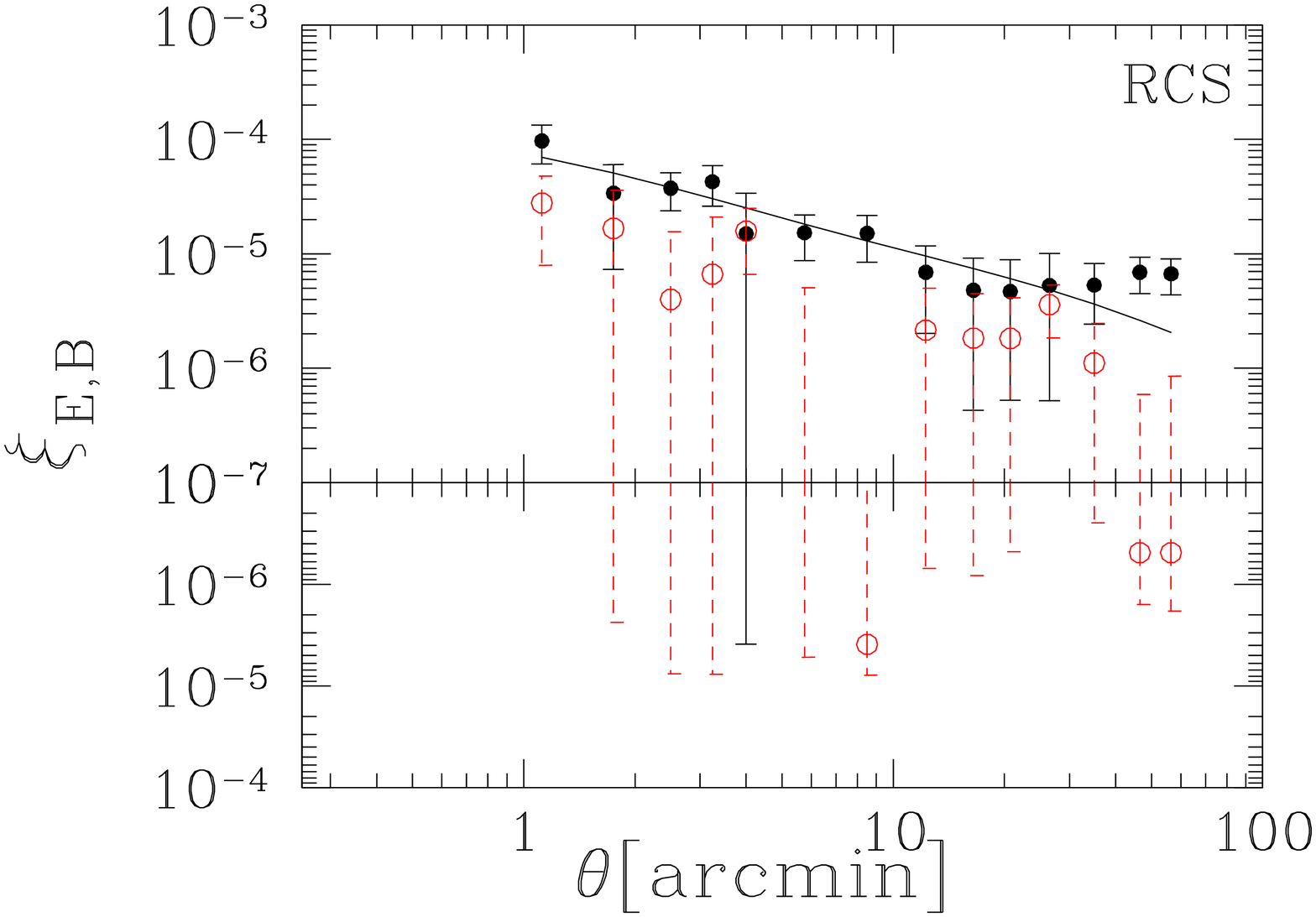}
\includegraphics[scale=0.33]{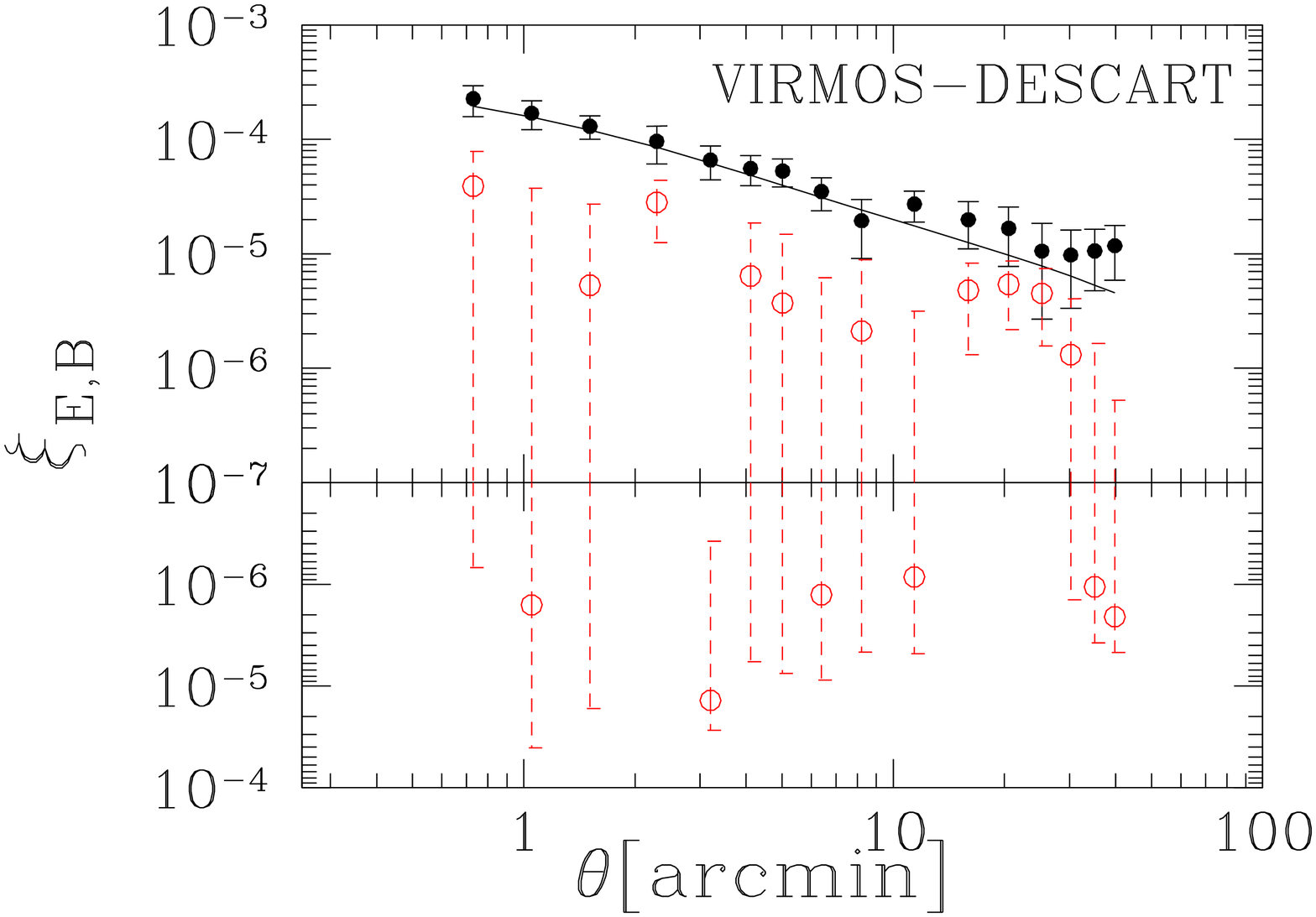}
}
}
\caption{\label{fig:shear} $E$ and $B$ modes of the shear correlation function $\xi$ (filled and open points, respectively) as measured for each survey. Note that the $1\sigma$ errors on the $E$-modes include statistical noise, non-Gaussian sample variance (see \S6) and a systematic error given by the magnitude of the $B$-mode.  The $1\sigma$ error on the $B$-modes is statistical only. The results are presented on a log-log scale, despite the existence of negative $B$-modes. We have therefore collapsed the infinite space between $10^{-7}$ and zero, and plotted negative values on a separate log scale mirrored on $10^{-7}$. Hence all values on the lower portion of the graph are negative, their absolute value is given by the scaling of the graph. Note that this choice of scaling exaggerates any discrepancies. The solid lines show the best fit $\Lambda$CDM model for $\Omega_m=0.24$, $h=0.72$, $\Gamma = h\Omega_m$, $\sigma_8$ given in Table~\ref{sigma8s}, and $n(z)$ given in Table~\ref{redshiftfits}. The latter two being chosen for the case of the high confidence redshift calibration sample, an $n(z)$ modeled by Eq.(\ref{brainerd}), and the non-linear power spectrum estimated by \citet{2003MNRAS.341.1311S}.}
\end{center}
\end{figure*}

Lensing shear has been measured for each survey using the common KSB galaxy shape measurement method \citep{1995ApJ...449..460K,1997ApJ...475...20L,1998ApJ...504..636H}.  The practical application of this method to the four surveys differs slightly, as tested by the Shear TEsting Programme; STEP \citep{2007MNRAS.376...13M,2006MNRAS.368.1323H}, resulting in a small calibration bias in the measurement of the shear. \citet{2006MNRAS.368.1323H} define both a calibration bias ($m$) and an offset of the shear ($c$),
\begin{equation}
 \gamma_{\mathrm{meas}} - \gamma_{\mathrm{true}} = m \gamma_{\mathrm{true}} + c, \label{bias}
\end{equation}
\noindent where $\gamma_{\mathrm{meas}}$ is the measured shear, $\gamma_{\mathrm{true}}$ is the true shear signal, and $m$ and $c$ are determined from an analysis of simulated data. The value of $c$ is highly dependent on the strength of the PSF distribution, therefore the value determined through comparisons with simulated data can not be easily applied to real data whose PSF strength varies across the image. Since an offset in the shear will appear as residual B-modes, we take $c=0$, and focus on the more significant and otherwise unaccounted for shear bias ($m$). Eq.(\ref{bias}) then simplifies to
\begin{equation} 
\gamma_{\mathrm{true}} = \gamma_{\mathrm{meas}} (m+1)^{-1}. 
\end{equation}
The calibration bias for the CFHTLS-Wide, RCS and VIRMOS-DESCART (HH analysis in \citet{2007MNRAS.376...13M}) is $m = -0.0017 \pm 0.0088$, and $m = 0.038 \pm 0.026$ for GaBoDS (MH analysis in \citet{2007MNRAS.376...13M}).

We measure the shear correlation functions $\xi_E(\theta)$ and $\xi_B(\theta)$ (Eq.~\ref{eqn:xieb}) for each survey, shown in Figure~\ref{fig:shear}. As discussed in \S\ref{sec:theory}, these statistics are calibrated using the measured aperture mass $B$-mode $\langle M_{\rm ap}^2\rangle_\perp$. The $1\sigma$ errors on $\xi_E$ include statistical noise, computed as described in \citet{2002A&A...396....1S}, non-Gaussian sample variance (see \S6 for details), and a systematic error added in quadrature.  The systematic error for each angular scale $\theta$ is given by the magnitude of $\xi_B(\theta)$.  The $1\sigma$ errors on $\xi_B$ are statistical only. We correct for the calibration bias by scaling $\xi_E$ by $(1 + m)^{-2}$, the power of negative 2 arising from the fact that $\xi_E$ is a second order shear statistic. When estimating parameter constraints in \S6.2 we marginalise over the error on the calibration bias, which expresses the range of admissible scalings of $\xi_E$. Tables containing the measured correlation functions and their corresponding covariance matrices are included as Supplementary Material to the online version of this article.

\section{Redshift Distribution}

\begin{table*}
\caption{The best fit model parameters for the redshift distribution, given by either Eq.(\ref{brainerd}) or Eq.(\ref{heymans}). Models are fit using the full calibration sample of \protect\citet{Ilbert..0603217} $0.0 \leq z_p \leq 4.0$, and using only the high confidence region $0.2 \leq z_p \leq 1.5$. $\chi^2_{\nu}$ is the reduced $\chi^2$ statistic, $\langle z \rangle$, and $z_\mathrm{m}$ are the average and median redshift respectively, calculated from the model on the range $0.0 \leq z \leq 4.0$.}
\label{redshiftfits}
\begin{center}
\begin{tabular}{lllllllllllllll}
\hline
\hline
    &    & \multicolumn{6}{c}{Eq.(\ref{brainerd})} &  \multicolumn{6}{c}{Eq.(\ref{heymans})}\\
\hline
    &    & $\alpha$ & $\beta$ & $z_0$ & $\langle z \rangle$ & $z_\mathrm{m}$ & $\chi^2_{\nu}$ & $a$ & $b$ & $c$ & $N$ & $\langle z \rangle$ & $z_\mathrm{m}$ & $\chi^2_{\nu}$\\
\multirow{4}{*}{$0.2 \leq z_p \leq 1.5$} 
& CFHTLS-Wide   & 0.836 & 3.425 & 1.171 & 0.802 & 0.788 & 1.63 & 0.723 & 6.772 & 2.282 & 2.860 & 0.848 & 0.812 & 1.65 \\
& GaBoDS & 0.700 & 3.186 & 1.170 & 0.784 & 0.760 & 1.96 & 0.571 & 6.429 & 2.273 & 2.645 & 0.827 & 0.784 & 2.80 \\
& RCS    & 0.787 & 3.436 & 1.157 & 0.781 & 0.764 & 1.41 & 0.674 & 6.800 & 2.095 & 2.642 & 0.823 & 0.788 & 1.60 \\
& VIRMOS-DESCART & 0.637 & 4.505 & 1.322 & 0.823 & 0.820 & 1.48 & 0.566 & 7.920 & 6.107 & 6.266 & 0.859 & 0.844 & 2.06 \\
\hline
\multirow{4}{*}{$0.0 \leq z_p \leq 4.0$} 
& CFHTLS-Wide   & 1.197 & 1.193 & 0.555 & 0.894 & 0.788 & 8.94 & 0.740 & 4.563 & 1.089 & 1.440 & 0.945 & 0.828 & 2.41 \\
& GaBoDS & 1.360 & 0.937 & 0.347 & 0.938 & 0.800 & 7.65 & 0.748 & 3.932 & 0.800 & 1.116 & 0.993 & 0.832 & 2.40 \\
& RCS    & 1.423 & 1.032 & 0.391 & 0.888 & 0.772 & 5.67 & 0.819 & 4.418 & 0.800 & 1.201 & 0.939 & 0.808 & 1.84 \\
& VIRMOS-DESCART & 1.045 & 1.445 & 0.767 & 0.909 & 0.816 & 9.95 & 0.703 & 5.000 & 1.763 & 2.042 & 0.960 & 0.864 & 2.40 \\
\hline
\hline
\end{tabular}
\end{center}
\end{table*}

The measured weak lensing signal depends on the redshift distribution of the sources, as seen from Eq.(\ref{pofkappa}). Past weak lensing studies (e.g., \citet{2006ApJ...647..116H}; \citet{2006AA...452...51S}; \citet{2005MNRAS.359.1277M}; \citet{2005AA...429...75V}) have used the Hubble Deep Field (HDF) photometric redshifts to estimate the shape of the redshift distribution. Spanning only 5.3 arcmin$^2$, the HDF suffers from sample variance as described in \citet{Waerbeke..0603696}, where it is also suggested that the HDF fields may be subject to a selection bias.  This sample variance of the measured redshift distribution adds an additional error to the weak lensing analysis that is not typically accounted for.

In this study we use the largest deep photometric redshift catalogue in existence, from \citet{Ilbert..0603217}, who have estimated redshifts on the four Deep fields of the T0003 CFHTLS release.  The redshift catalogue is publicly available at $\it{terapix.iap.fr}$. The full photometric catalogue contains 522286 objects, covering an effective area of 3.2 deg$^2$. A set of 3241 spectroscopic redshifts with $0 \leq z \leq 5$ from the VIRMOS VLT Deep Survey (VVDS) were used as a calibration and training set for the photometric redshifts. The resulting photometric redshifts have an accuracy of $\sigma_{(z_{phot} - z_{spec})/(1+z_\mathrm{spec})} = 0.043$ for i'$_{\mathrm{AB}}$ = 22.5 - 24, with a fraction of catastrophic errors of 5.4$\pct$. 
\citet{Ilbert..0603217} demonstrate that their derived redshifts work best in the range $0.2 \leq z \leq 1.5$, having a fraction of catastrophic errors of $\sim5\pct$ in this range. The fraction of catastrophic errors increases dramatically at $z < 0.2$ and $1.5 < z < 3$ reaching $\sim40\pct$ and $\sim70\pct$ respectively. This is explained by a degeneracy between galaxies at $z_{\mathrm{spec}} < 0.4$ and $1.5 < z_{\mathrm{phot}} < 3$ due to a mismatch between the Balmer break and the intergalactic Lyman-alpha forest depression. 

\citet{Waerbeke..0603696} estimate the expected sampling error on the average redshift for such a photometric redshift sample to be $\sim3\pct / \sqrt{4} = 1.5\pct$, where the factor of $\sqrt{4}$ comes from the four independent CFHTLS-Deep fields, a great improvement over the $\sim10\pct$ error expected for the HDF sample.

\subsection{Magnitude Conversions}
In order to estimate the redshift distributions of the surveys, we calibrate the magnitude distribution of the photometric redshift sample (hence forth the $z_\mathrm{p}$ sample) to that of a given survey by converting the CFHTLS filter set $u^*g'r'i'z'$ to $I_{\mathrm{AB}}$ magnitudes for VIRMOS-DESCART and $R_{\mathrm{VEGA}}$ magnitudes for RCS and GaBoDS.  We employ the linear relationships between different filter bands given by \citet{2007AJ....133..734B}:

\begin{eqnarray}
I_{\mathrm{AB}}&=&i' - 0.0647 - 0.7177 \left[ (i'-z') - 0.2083 \right], \\
R_{\mathrm{VEGA}}&=&r' - 0.0576 - 0.3718 \left[ (r'-i') - 0.2589 \right] - 0.21.\nonumber
\end{eqnarray}

\noindent These conversions were estimated by fitting spectral energy distribution templates to data from the Sloan Digital Sky Survey.  They are considered to be accurate to 0.05 mag or better, resulting in an error on our estimated median redshifts of at most $1\%$, which is small compared to the total error budget on the estimated cosmological parameters.

\subsection{Modeling the redshift distribution $n(z)$}
The galaxy weights used in each survey's lensing analysis result in a weighted source redshift distribution.  To estimate the effective $n(z)$ of each survey we draw galaxies at random from the $z_p$ sample, using the method described in \S6.5.1 of \citet{2003psa..book.....W} to reproduce the shape of the weighted magnitude distribution of each survey.
A Monte Carlo (bootstrap) approach is taken to account for the errors in the photometric redshifts, as well as the statistical variations expected from drawing a random sample of galaxies from the $z_\mathrm{p}$ sample to create the redshift distribution.  The process of randomly selecting galaxies from the $z_\mathrm{p}$ sample, and therefore estimating the redshift distribution of the survey, is repeated 1000 times. Each redshift that is selected is drawn from its probability distribution defined by the $\pm1$ and $\pm3 \sigma$ errors (since these are the error bounds given in \citet{Ilbert..0603217}'s catalogue). The sampling of each redshift is done such that a uniformly random value within the $1\sigma$ error is selected 68$\pct$ of the time, and a uniformly random value within the three sigma error (but exterior to the $1\sigma$ error) is selected 32$\pct$ of the time. The redshift distribution is then defined as the average of the 1000 constructions, and an average covariance matrix of the redshift bins is calculated. 

At this point sample variance and Poisson noise are added to the diagonal elements of the covariance matrix, following \citet{Waerbeke..0603696}.  They provide a scaling relation between cosmic (sample) variance noise and Poisson noise where $\sigma_{\mathrm{sample}}/\sigma_{\mathrm{Poisson}}$ is given as a function of redshift for different sized calibration samples. We take the curve for a 1 deg$^2$ survey, the size of a single CFHTLS-Deep field, and divide by $\sqrt{4}$ since there are four independent 1 deg$^2$ patches of sky in the photometric redshift sample.

\begin{figure*}
\begin{center}
\hbox{
\includegraphics[scale=0.36]{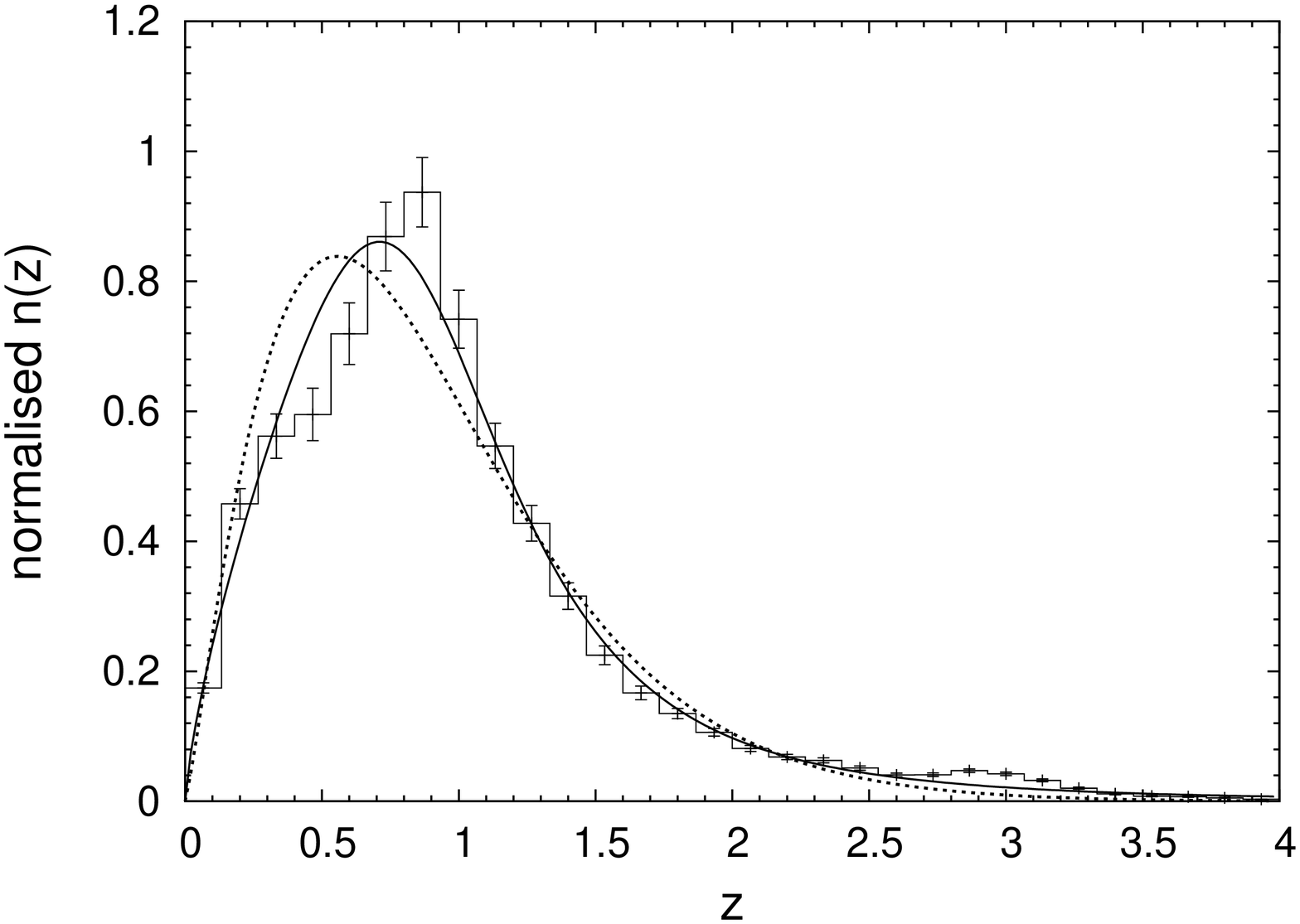}
\includegraphics[scale=0.36]{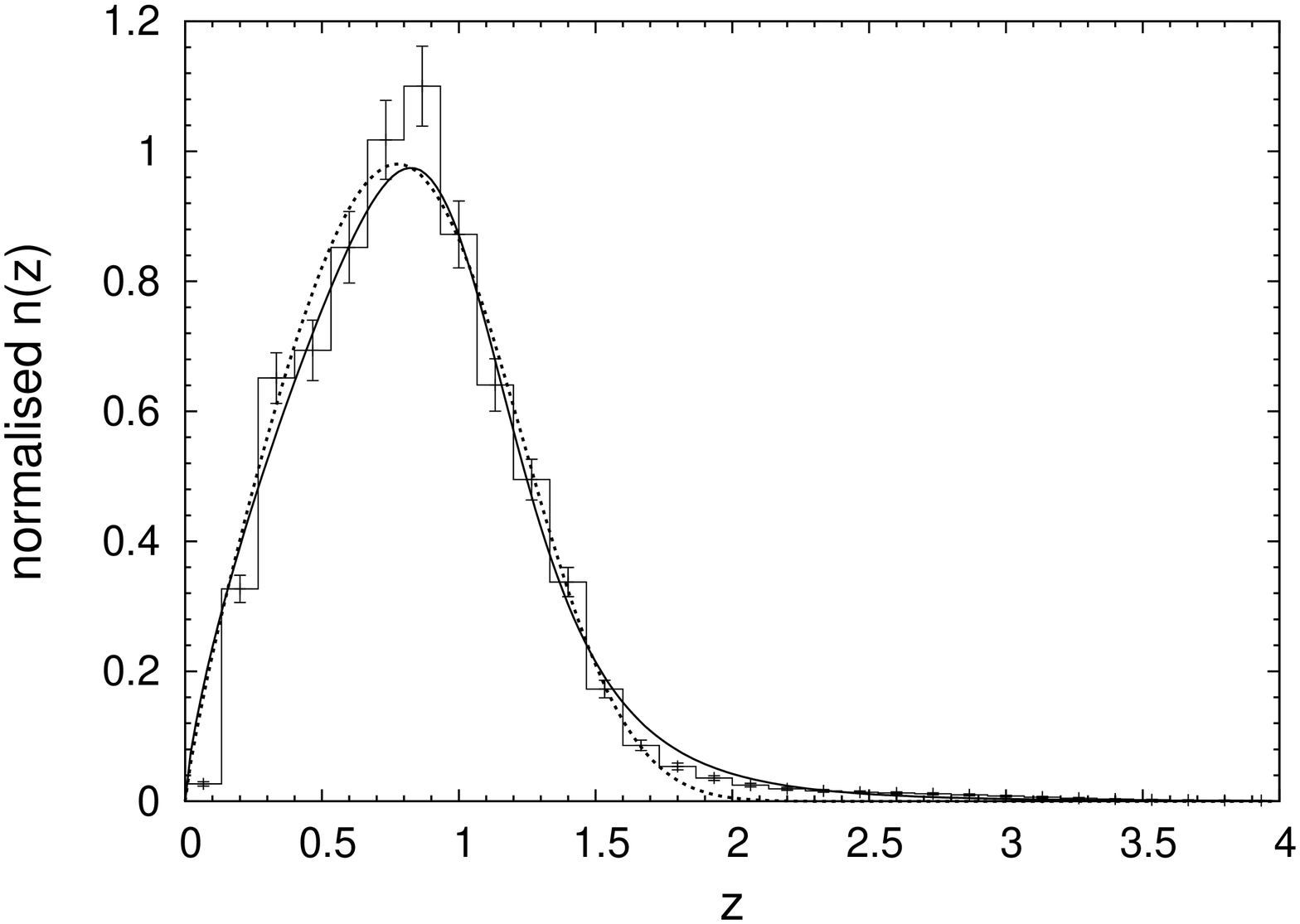}
}
\caption{\label{widedist} Normalised redshift distribution for the CFHTLS-Wide survey, given by the histogram, where the error bars include Poisson noise and sample variance of the photometric redshift sample. The dashed curve shows the best fit for Eq.(\ref{brainerd}), and the solid curve for Eq.(\ref{heymans}). \textbf{Left:} Distribution obtained if all photometric redshifts are used $0.0 \leq z \leq 4.0$, and $\chi^2$ is calculated between $0.0 \leq z \leq 2.5$. \textbf{Right:} Distribution obtained if only the high confidence redshifts are used $0.2 \leq z \leq 1.5$, and $\chi^2$ is calculated on this range. The existence of counts for $z > 1.5$ is a result of drawing the redshifts from their full probability distributions.}
\end{center}
\end{figure*}

Based on the photometric redshift sample of \citet{Ilbert..0603217} we consider two redshift ranges, the full range of redshifts in the photometric catalogues $0.0 \leq z \leq 4.0$, and the high confidence range $0.2 \leq z \leq 1.5$.  The goal is to assess to what extent this will affect the redshift distribution, and --in turn-- the derived parameter constraints. 

The shape of the normalised redshift distribution is often assumed to take the following form:
\begin{equation}
n(z)=\frac {\beta}{z_0\; \Gamma \Big(\frac{1+\alpha}{\beta}\Big)}
\Big(\frac{z}{z_0}\Big)^\alpha\;\exp\,\Big[{-\Big(\frac{z}{z_0}\Big)^\beta}\Big],
\label{brainerd}
\end{equation}
\noindent where $\alpha$, $\beta$, and $z_0$ are free parameters. If the full range of photometric redshifts is used the shape of the redshift distribution is poorly fit by Eq.(\ref{brainerd}), this is a result of the function's exponential drop off which can not accommodate the number of high redshift galaxies in the tail of the distribution (see Fig.(\ref{widedist})). We therefore adopt a new function in an attempt to better fit the normalised redshift distribution,
\begin{equation}
 n(z)=N\frac{z^a}{z^b + c},
\label{heymans}
\end{equation}
\noindent where $a$, $b$, and $c$ are free parameters, and $N$ is a normalising factor,
\begin{equation}
 N=\Big(\int_0^\infty{dz'\frac{z'^a}{z'^b + c}}\Big)^{-1}.
\end{equation}

The best fit model is determined by minimising the generalised chi square statistic:
\begin{equation}
 \chi^2 = (d_i-m_i)\Cg^{-1}(d_i-m_i)^T,
\end{equation}
where $\Cg$ is the covariance matrix of the binned redshift distribution as determined from the 1000 Monte Carlo constructions, $d_i$ is the number count of galaxies in the $i^{\rm th}$ bin from the average of the 1000 distributions, and $m_i$ is the number count of galaxies at the center of the $i^{\rm th}$ bin for a given model distribution. For each survey we determine the best fit for both Eq.(\ref{brainerd}) and Eq.(\ref{heymans}). In either case we consider both the full range of redshifts and the high confidence range, the results are presented in Table~\ref{redshiftfits}. 

Figure~\ref{widedist} shows the best fit models to the CFHTLS-Wide survey, when the full range of photometric redshifts is considered (left panel), Eq.(\ref{heymans}) clearly fits the distribution better than Eq.(\ref{brainerd}), having reduced $\chi^2$ statistics of 2.41 and 8.94 respectively. When we consider only the high confidence redshifts (right panel of Figure~\ref{widedist}) both functions fit equally well having reduced $\chi^2$ statistics of 1.65 and 1.63 for Eq.(\ref{heymans}) and Eq.(\ref{brainerd}) respectively. Note that the redshift distribution as determined from the data (histograms in Figure \ref{widedist}) is non-zero at $z > 1.5$ for the high confidence photometric redshifts ($0.2 \leq z_\mathrm{p} \leq 1.5$) because of the Monte Carlo sampling of the redshifts from their probability distributions.

\section{Parameter estimation}
\subsection{Maximum likelihood method}
We investigate a six dimensional parameter space consisting of the mean matter density $\Omega_\mathrm{m}$, the normalisation of the matter power spectrum $\sigma_\mathrm{8}$, the Hubble parameter $h$, and $n(z)$ the redshift distribution parametrised by either $\alpha$, $\beta$, and $z_0$ (Eq.\ref{brainerd}) or a, b, and c (Eq.\ref{heymans}). A flat cosmology ($\Omega_\mathrm{m} + \Omega_{\Lambda} = 1$) is assumed throughout, and the shape parameter is given by $\Gamma = \Omega_\mathrm{m}\,h$. The default priors are taken to be $\Omega_\mathrm{m} \in [0.1,1], \sigma_\mathrm{8} \in [0.5,1.2]$, and $h \in [0.64,0.8]$ with the latter in agreement with the findings of the HST key project \citep{2001ApJ...553...47F}. The priors on the redshift distribution were arrived at using a Monte Carlo technique. This is necessary since the three parameters of either Eq.(\ref{brainerd}) or Eq.(\ref{heymans}) are very degenerate, hence simply finding the 2 or 3 $\sigma$ levels of one parameter while keeping the other two fixed at their best fit values does not fairly represent the probability distribution of the redshift parameters.  The method used ensures a sampling of parameter triplets whose number count follow the 3-D probability distribution; that is 68$\pct$ lie within the 1$\sigma$ volume, 97$\pct$ within the 2$\sigma$ volume, etc. We find 100 such parameter trios and use them as the prior on the redshift distribution, therefore this is a Gaussian prior on $n(z)$. 

Given the data vector $\xig$, which is the shear correlation function ($\xi_E$ of Eq.\ref{eqn:xieb}) as a function of scale, and the model prediction $\mg(\Omega_\mathrm{m},\sigma_\mathrm{8},h,n(z))$ the likelihood function of the data is given by

\begin{equation}
{\cal L}={1\over \sqrt{(2\pi)^n|\Cg|}} \exp\left[-\frac{1}{2}(\xig-\mg)\Cg^{-1}(\xig-\mg)^T\right], \label{likelihood}
\end{equation}
where $n$ is the number of angular scale bins and $\Cg$ is the $n\times n$ covariance matrix. The shear covariance matrix can be expressed as
\begin{equation}
C_{ij}=\langle (\xi_i-\mu_i)(\xi_j-\mu_j)\rangle,
\end{equation}
where $\mu_i$ is the mean of the shear $\xi_i$ at scale $i$, and angular brackets denote the average over many independent patches of sky. To obtain a reasonable estimate of the covariance matrix for a given set of data one needs many independent fields, this is not the case for either the CFHTLS-Wide or VIRMOS-DESCART surveys. 

We opt to take a consistent approach for all 4 data sets by decomposing the shear covariance matrix as $\Cg=\Cg_n+\Cg_B+\Cg_s$, where $\Cg_n$ is the statistical noise, $\Cg_B$ is the absolute value of the residual $B$-mode and $\Cg_s$ is the sample variance covariance matrix. $\Cg_n$ can be measured directly from the data, it represents the statistical noise inherent in a finite data set. $\Cg_B$ is diagonal and represents the addition of the $B$-mode in quadrature to the uncertainty, this provides a conservative limit to how well the lensing signal can be determined.

\begin{table}
\caption{Data used for each survey to calculate the Gaussian contribution to the sample variance covariance matrix. $A_{\mathrm{eff}}$ is the effective area, $n_{\mathrm{eff}}$ the effective galaxy number density, and $\sigma_e$ the intrinsic ellipticity dispersion.\label{cosvar}}
\label{covarmatdata}
\begin{center}
\begin{tabular}{lccc}
\hline
\hline
& $A_{\mathrm{eff}}$ & $n_{\mathrm{eff}}$ & $\sigma_e$\\
& $(\mathrm{deg}^2)$ & $(\mathrm{arcmin}^{-2})$ & \\
\hline
CFHTLS-Wide & 22 & 12 & 0.47 \\
RCS & 53 & 8 & 0.44 \\
VIRMOS-DESCART & 8.5 & 15 & 0.44 \\
GaBoDS & 13 & 12.5 & 0.50\\
\hline
\hline
\end{tabular}
\end{center}
\end{table}

For Gaussian sample variance the matrix $\Cg_s$ can be computed according to \citet{2002A&A...396....1S}, assuming an effective survey area $A_{\mathrm{eff}}$, an effective number density of galaxies $n_{\mathrm{eff}}$, and an intrinsic ellipticity dispersion $\sigma_e$ (see Table~\ref{cosvar}). However, \citet{2002A&A...396....1S} assume Gaussian statistics for the fourth order moment of the shear correlation function --a necessary simplification to achieve an analytic form for the sample variance covariance matrix. We use the calibration presented by \citet{Semboloni..0606648}, estimated from ray tracing simulations, to account for non-Gaussianities. Their work focuses on the following quantity:
\begin{equation}
{\cal F}(\vartheta_1 ,\vartheta_2 )= {\rm C_s^{\mathrm{meas}}(\xi_+;\vartheta_1,\vartheta_2)\over
C_s^{\mathrm{Gaus}}(\xi_+;\vartheta_1,\vartheta_2)}, \label{fudge}
\end{equation}
where ${\cal F}(\vartheta_1 ,\vartheta_2 )$ is the ratio of the sample variance covariance measured from N-body simulations ($\rm C_s^{\mathrm{meas}}(\xi_+;\vartheta_1,\vartheta_2)$) to that expected from Gaussian effects alone ($\rm C_s^{\mathrm{Gaus}}(\xi_+;\vartheta_1,\vartheta_2)$). It is found that ${\cal F}(\vartheta_1 ,\vartheta_2 )$ increases significantly above unity at scales smaller than $\sim10$ arcminutes, increasing with decreasing scale, it reaches an order of magnitude by $\sim2$ arcminutes. The parametrised fit as a function of mean source redshift is given by,
\begin{eqnarray}
{\cal F}(\vartheta_1 ,\vartheta_2 )&=&{p_1(z)\over \left[ \vartheta_1^2 \vartheta_2^2\right] ^{p_2(z)}}\label{pparam} \, , \\
p_1(z)&=&{16.9\over z  ^{0.95}}-2.19\nonumber \, ,\\
p_2(z)&=&1.62\: z^{-0.68}\: {\rm exp}( -z ^ {-0.68})-0.03. \nonumber 
\end{eqnarray}

A fiducial model is required to calculate the Gaussian covariance matrix, it is taken as $\Omega_\mathrm{m}=0.3$, $\Lambda=0.7$, $\sigma_\mathrm{8}=0.8$, $h=0.72$, and the best fit $n(z)$ model (see Table~\ref{redshiftfits}). We then use the above prescription to account for non-Gaussianities, increasing $\Cg_s$ everywhere ${\cal F}$ is above unity. For each survey the total covariance matrix ($\Cg = \Cg_n + \Cg_s + \Cg_B$) is included as Supplementary Matrial to the online version of this article.
%

To test this method we compare our analytic covariance matrix for GaBoDS with that found by measuring it from the data \citep{Hetterscheidt..0606571}. Since GaBoDS images 52 independent fields it is possible to obtain an estimate of $\Cg$ directly from the data. The contribution from the $B$-modes ($\Cg_B$) is not added to our analytic estimate in this case, since its inclusion is meant as a conservative estimate of the systematic errors.  We find a median percent difference along the diagonal of $\sim15\pct$, which agrees well with the accuracy obtained for simulated data \citep{Semboloni..0606648}. 

\begin{figure}
\begin{center}
\includegraphics[scale=0.45]{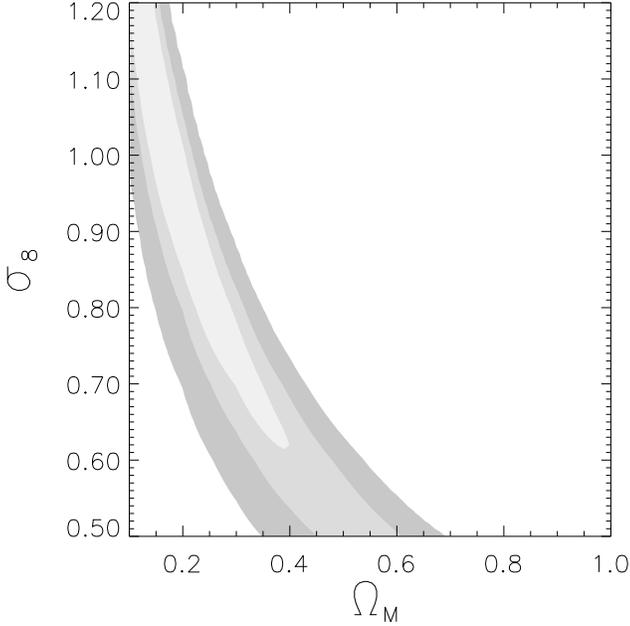}
\caption{\label{Fig:combinedconst} Joint constraints on $\sigma_\mathrm{8}$ and $\Omega_\mathrm{m}$ from the 100 deg$^2$ weak lensing survey assuming a flat $\Lambda$CDM cosmology and adopting the non-linear matter power spectrum of \protect\citet{2003MNRAS.341.1311S}. The redshift distribution is estimated from the high confidence photometric redshift catalogue \protect\citep{Ilbert..0603217}, and modeled with the standard functional form given by Eq.(\ref{brainerd}). The contours depict the 0.68, 0.95, and $0.99\pct$ confidence levels. The models are marginalised, over $h=0.72\pm0.08$, shear calibration bias (see \S4) with uniform priors,  and the redshift distribution with Gaussian priors (see \S6.1). Similar results are found for all other cases, as listed in Table~\ref{sigma8s}.}
\end{center}
\end{figure}
\begin{figure}
\begin{center}
\vbox{
\hbox{
\includegraphics[scale=0.22]{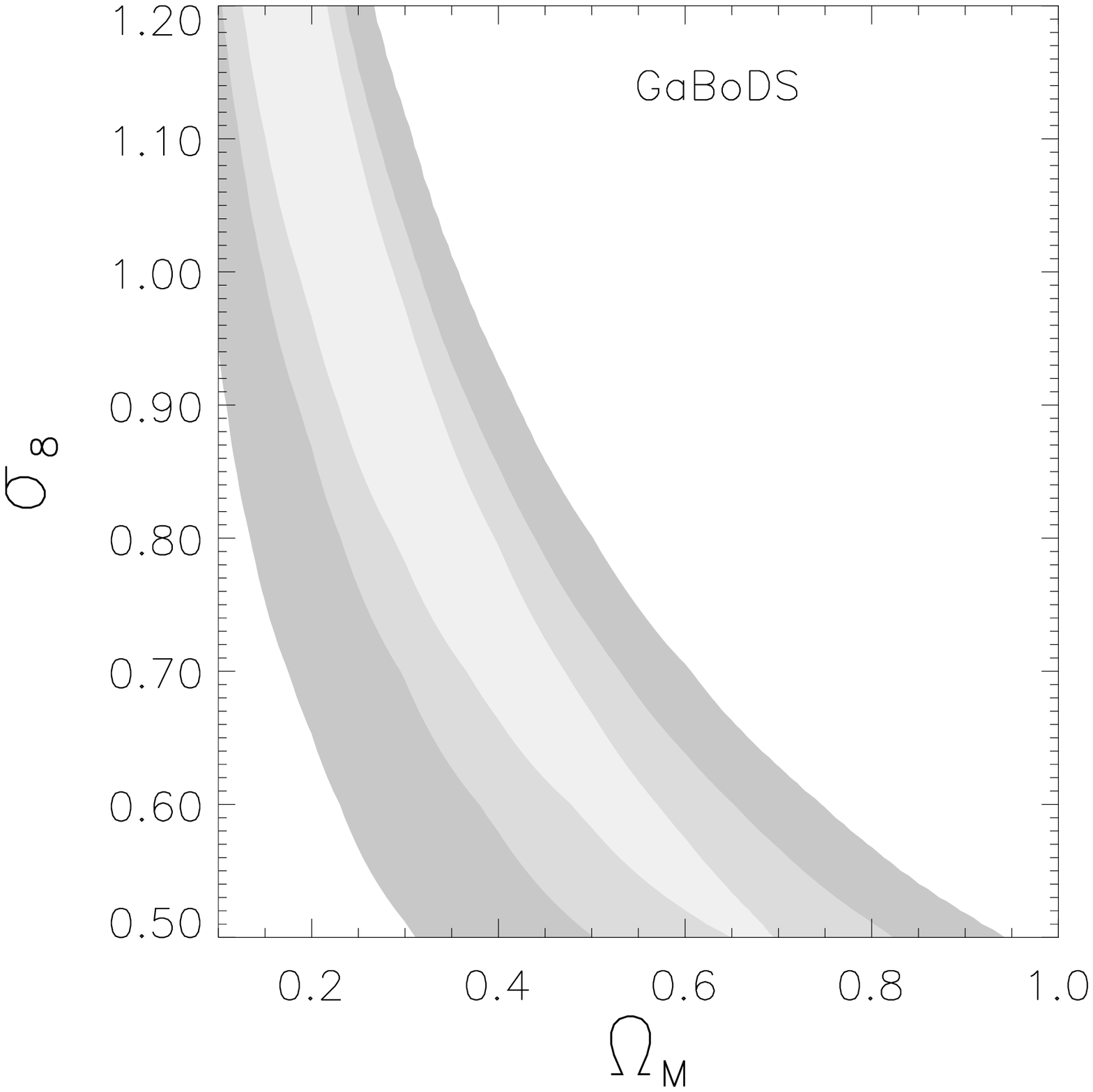}
\includegraphics[scale=0.22]{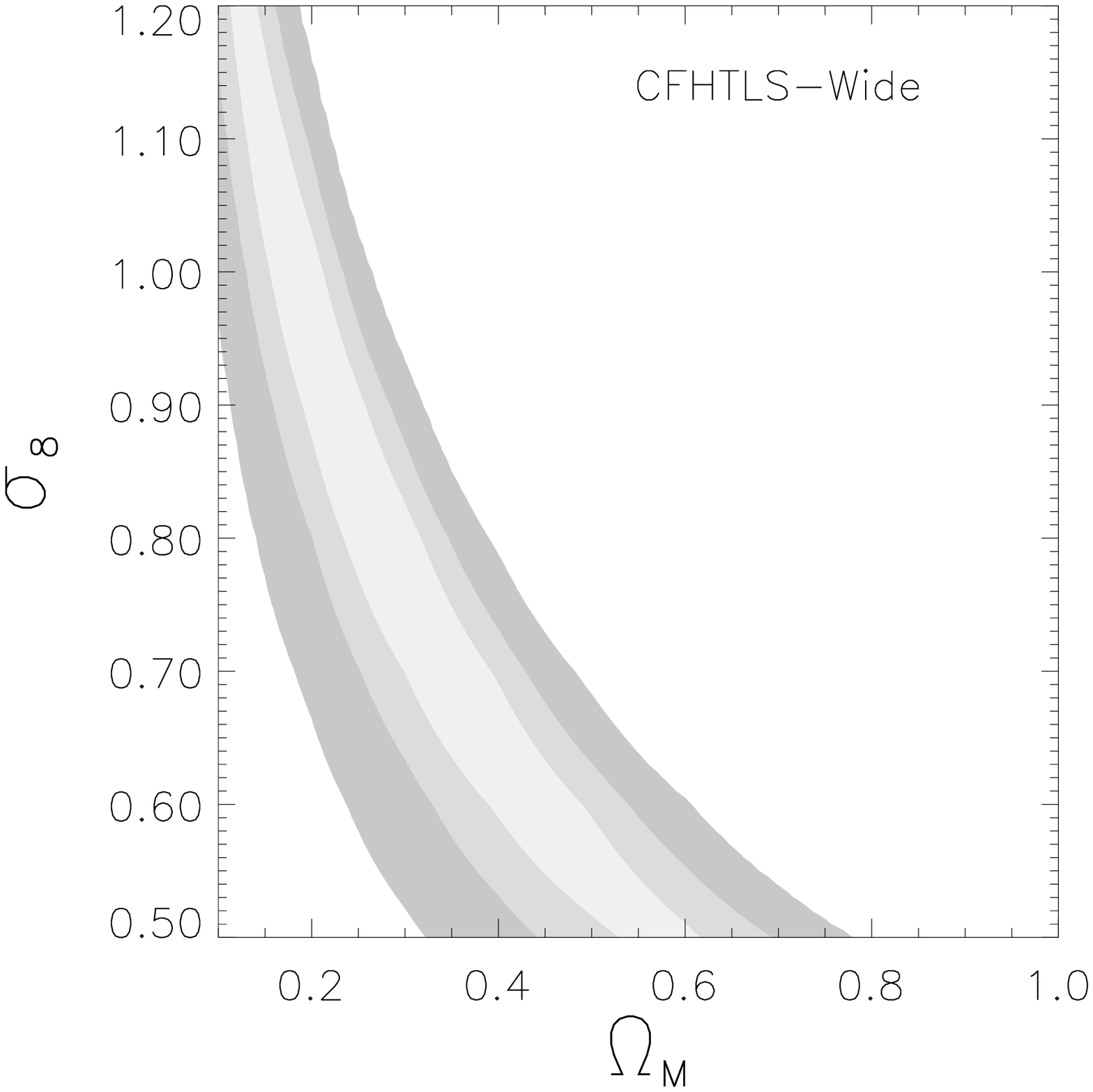}
}
\hbox{
\includegraphics[scale=0.22]{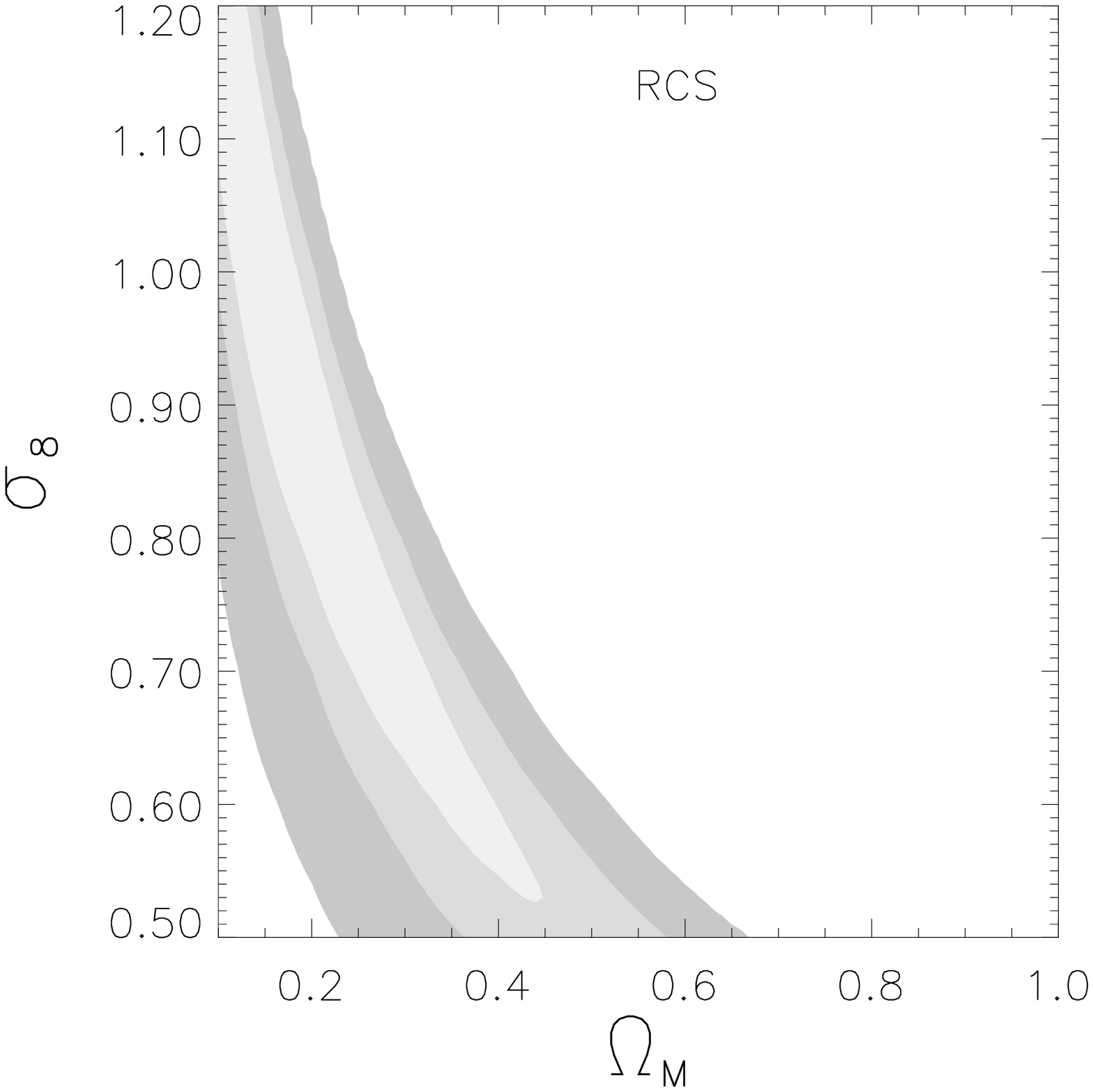}
\includegraphics[scale=0.22]{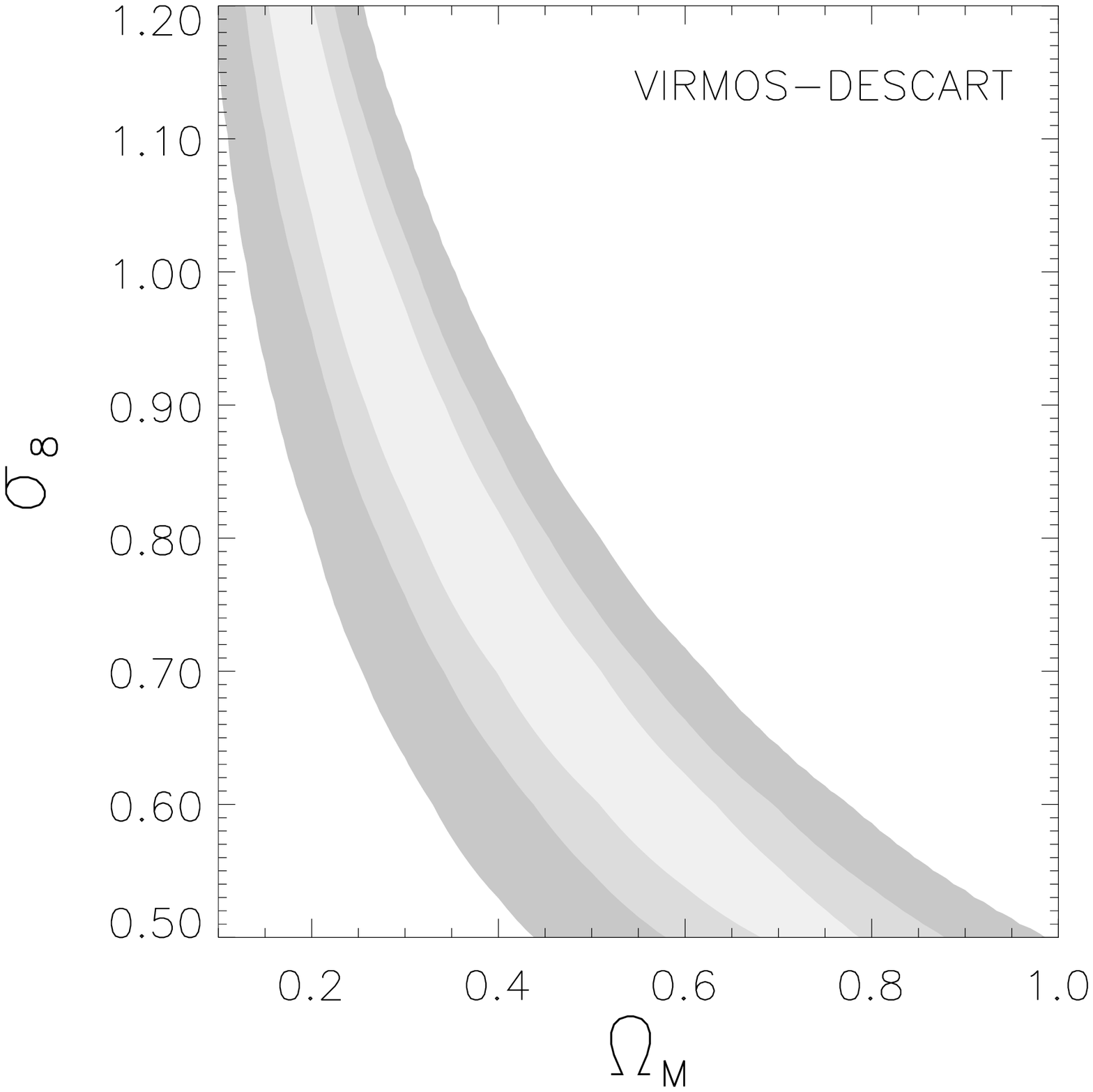}
}
}
\caption{\label{paramconst} Joint constraints on $\sigma_\mathrm{8}$ and $\Omega_\mathrm{m}$ for each survey assuming a flat $\Lambda$CDM cosmology and adopting the non-linear matter power spectrum of \protect\citet{2003MNRAS.341.1311S}. The redshift distribution is estimated from the high confidence photometric redshift catalogue \protect\citep{Ilbert..0603217}, and modeled with the standard functional form given by Eq.(\ref{brainerd}). The contours depict the 0.68, 0.95, and $0.99\pct$ confidence levels. The models are marginalised, over $h=0.72\pm0.08$, shear calibration bias (see \S4) with uniform priors,  and the redshift distribution with Gaussian priors (see \S6.1). Similar results are found for all other cases, as listed in Table~\ref{sigma8s}.}
\end{center}
\end{figure}
\subsection{Joint constraints on $\sigma_\mathrm{8}$ and $\Omega_\mathrm{m}$}
\begin{table*}
\caption{Joint constraints on the amplitude of the matter power spectrum $\sigma_\mathrm{8}$, and the matter energy density $\Omega_\mathrm{m}$, parametrised as $\sigma_\mathrm{8}(\frac{\Omega_\mathrm{m}}{0.24})^{\alpha}=\beta$. The errors are $1\sigma$ limits, calculated for a hard prior of $\Omega_\mathrm{m}=0.24$.}  
\label{sigma8s}
\begin{center}
\begin{tabular}{llllllllll}
\hline
\hline
    &    & \multicolumn{4}{c}{$0.0 \leq z_\mathrm{p} \leq 4.0$} &  \multicolumn{4}{c}{$0.2 \leq z_\mathrm{p} \leq 1.5$} \\
\hline
&    & \multicolumn{2}{c}{Eq.(\ref{brainerd})} & \multicolumn{2}{c}{Eq.(\ref{heymans})} & \multicolumn{2}{c}{Eq.(\ref{brainerd})} & \multicolumn{2}{c}{Eq.(\ref{heymans})} \\
&    & \multicolumn{1}{c}{$\beta$} & \multicolumn{1}{c}{$\alpha$} & \multicolumn{1}{c}{$\beta$} & \multicolumn{1}{c}{$\alpha$} & \multicolumn{1}{c}{$\beta$} & \multicolumn{1}{c}{$\alpha$} & \multicolumn{1}{c}{$\beta$} & \multicolumn{1}{c}{$\alpha$}\\
\multirow{5}{*}{\citet{2003MNRAS.341.1311S}}
& CFHTLS-Wide & $0.84 \pm 0.06$ & $0.55$ & $0.81 \pm 0.07$ & $0.54$ & $0.86 \pm 0.06$ & $0.56$ & $0.84 \pm 0.06$ & $0.55$ \\
& GaBoDS & $0.93 \pm 0.08$ & $0.60$ & $0.89 \pm 0.09$ & $0.59$ & $1.01 \pm 0.09$ & $0.66$ & $0.98 \pm 0.10$ & $0.63$ \\
& RCS & $0.75 \pm 0.07$ & $0.55$ & $0.73 \pm 0.08$ & $0.55$ & $0.78 \pm 0.07$ & $0.57$ & $0.76 \pm 0.07$ & $0.56$ \\
& VIRMOS-DESCART & $0.99 \pm 0.08$ & $0.59$ & $0.95 \pm 0.07$ & $0.58$ & $1.02 \pm 0.07$ & $0.60$ & $1.00 \pm 0.08$ & $0.59$ \\
& Combined & $0.80 \pm 0.05$ & $0.57$ & $0.77 \pm 0.05$ & $0.56$ & $0.84 \pm 0.05$ & $0.59$ & $0.82 \pm 0.06$ & $0.58$ \\
\hline
\hline
\multirow{5}{*}{\citet{1996MNRAS.280L..19P}}
& CFHTLS-Wide & $0.87 \pm 0.07$ & $0.57$ & $0.83 \pm 0.06$ & $0.56$ & $0.89 \pm 0.06$ & $0.58$ & $0.87 \pm 0.06$ & $0.57$ \\
& GaBoDS & $0.96 \pm 0.09$ & $0.62$ & $0.93 \pm 0.08$ & $0.60$ & $1.04 \pm 0.09$ & $0.66$ & $1.01 \pm 0.09$ & $0.64$ \\
& RCS & $0.77 \pm 0.07$ & $0.57$ & $0.74 \pm 0.07$ & $0.56$ & $0.81 \pm 0.08$ & $0.58$ & $0.79 \pm 0.08$ & $0.58$ \\
& VIRMOS-DESCART & $1.02 \pm 0.07$ & $0.60$ & $0.98 \pm 0.07$ & $0.59$ & $1.05 \pm 0.07$ & $0.62$ & $1.03 \pm 0.08$ & $0.61$ \\
& Combined & $0.82 \pm 0.05$ & $0.58$ & $0.79 \pm 0.05$ & $0.57$ & $0.86 \pm 0.06$ & $0.60$ & $0.84 \pm 0.06$ & $0.59$ \\
\hline
\hline
\end{tabular}
\end{center}
\end{table*}
\begin{figure}
\begin{center}
\includegraphics[scale=0.35]{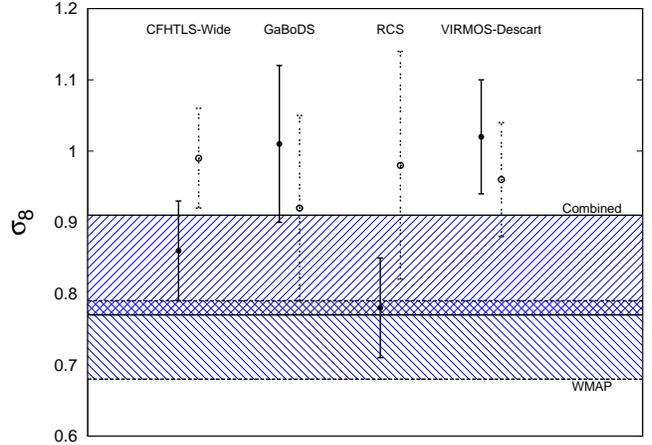}
\caption{\label{fig:compare0.24} Values for $\sigma_\mathrm{8}$ when $\Omega_\mathrm{m}$ is taken to be 0.24, filled circles (solid) give our results with $1\sigma$ error bars, open circles (dashed) show the results from previous analyses (Table~\ref{tab:summary}). Our results are given for the high confidence photometric redshift catalogue, using the functional form for $n(z)$ given by Eq.(\ref{brainerd}), and the \protect\citet{2003MNRAS.341.1311S} prescription for the non-linear power spectrum. The literature values use the \protect\citet{1996MNRAS.280L..19P} prescription for non-linear power, and are expected to be $\sim3\pct$ higher than would be the case for \protect\citet{2003MNRAS.341.1311S}. The forward slashed hashed region (enclosed by solid lines) shows the $1\sigma$ range allowed by our combined result, the back slashed hashed region (enclosed by dashed lines) shows the $1\sigma$ range given by the WMAP 3 year results.}
\end{center}
\end{figure}
For the combined survey we place joint constraints on $\sigma_\mathrm{8}$ and $\Omega_\mathrm{m}$ as shown in Figure~\ref{Fig:combinedconst}.  Fitting to the maximum likelihood region we find $\sigma_\mathrm{8} (\frac{\Omega_\mathrm{m}}{0.24})^{0.59} = 0.84\pm0.05$, where the quoted error is $1\sigma$ for a hard prior of $\Omega_\mathrm{m}=0.24$.  This result assumes a flat $\Lambda$CDM cosmology and adopts the non-linear matter power spectrum of \citet{2003MNRAS.341.1311S}.  The redshift distribution is estimated from the high confidence CFHTLS-Deep photometric redshifts using the standard $n(z)$ model given by Eq.(\ref{brainerd}).  Marginalisation was performed over $h\in[0.64,0.80]$ with flat priors, $n(z)$ with Gaussian priors as described in \S6.1, and a calibration bias of the shear signal with flat priors as discussed in \S4.  The corresponding constraints from each survey are presented in Figure~\ref{paramconst} and tabulated in Table~\ref{sigma8s}. Results in Table~\ref{sigma8s} are presented for two methods of calculating the non-linear power spectrum; \citet{1996MNRAS.280L..19P} and \citet{2003MNRAS.341.1311S}.  We find a difference of approximately $3\pct$ in the best-fit $\sigma_\mathrm{8}$ values, where results using \citet{2003MNRAS.341.1311S} are consistently the smaller of the two. The more recent results of \citet{2003MNRAS.341.1311S} are more accurate, and should be preferred over those of \citet{1996MNRAS.280L..19P}.  

The contribution to the error budget due to the conservative addition of the B-modes to the covariance matrix is small, amounting to at most 0.01 in the $1\sigma$ error bar for a hard prior of $\Omega_m=0.24$. The analysis was also performed having removed all scales where the B-modes are not consistent with zero (see Figure~\ref{fig:shear}). The resulting change in the best fit cosmology is small, shifting the best-fit $\sigma_8$ for an $\Omega_m$ of 0.24 by at most 0.03 (for CFHTLS-Wide), and on average by $\sim0.01$ across all four surveys. These changes are well within our error budget.

We present a comparison of the measured $\sigma_\mathrm{8}$ values with those previously published in Figure~\ref{fig:compare0.24}. The quoted $\sigma_\mathrm{8}$ values are for a vertical slice through $\sigma_\mathrm{8} - \Omega_\mathrm{m}$ space at an $\Omega_\mathrm{m}$ of 0.24, and all error bars denote the $1\sigma$ region from the joint constraint contours. Our error bars (filled circles) are typically smaller than those from the literature (open circles) mainly due to the improved estimate of the redshift distribution. Our updated result for each survey agrees with the previous analysis within the error bars, this remains true if the literature values are lowered by $\sim3\pct$ to account for the difference in methods used for the non-linear power specrum. Also plotted are the $1\sigma$ limits for the combined result and WMAP 3 year constraints (forward and back slashed hash regions respectively), our result is consistent with WMAP at the $1\sigma$ level.

In addition to our main analysis we have investigated the impact of using four different models for the redshift distribution.  These models are dependent on which functional form is used (Eq.(\ref{brainerd}) or Eq.(\ref{heymans})) and which range is used for the photometric redshift sample ($0.0 \leq z_p \leq 4.0$ or $0.2 \leq z_p \leq 1.5$). Table~\ref{sigma8s} gives the best fit joint constraints on $\sigma_\mathrm{8}$ and $\Omega_\mathrm{m}$, for each survey as well as the combined survey result.

Comparing the best fit $\sigma_\mathrm{8}$ values with the average redshifts listed in Table~\ref{redshiftfits}, we find, as expected, that higher redshift models result in lower values for $\sigma_\mathrm{8}$. Attempting to quantify this relation in terms of mean redshift fails. Changes in mean redshift are large ($\sim14\pct$) between the photometric samples, compared to $\sim6\pct$ between the two $n(z)$ models, however a $\sim5\pct$ change in $\sigma_\mathrm{8}$ is seen for both. The median redshift is a much better gauge, changing by $\sim4\pct$ between both photometric samples and $n(z)$ models. Precision cosmology at the $1\pct$ level will necessitate roughly the same level of precision of the median redshift. Future cosmic shear surveys will require thorough knowledge of the complete redshift distribution of the sources, in particular to what extent a high redshift tail exists.

\section{Discussion and Conclusion}
We have performed an analysis of the 100 square degree weak lensing survey that combines four of the largest weak lensing datasets in existence.  Our results provide the tightest weak lensing constraints on the amplitude of the matter power spectrum $\sigma_\mathrm{8}$ and matter density $\Omega_\mathrm{m}$ and a marked improvement on accuracy compared to previous results. Using the non-linear prediction of the cosmological power spectra given in \citet{2003MNRAS.341.1311S}, the high confidence region of the photometric redshift calibration sample, and Eq.(\ref{brainerd}) to model the redshift distribution, we find $\sigma_\mathrm{8}(\frac{\Omega_\mathrm{m}}{0.24})^{0.59}=0.84\pm0.05$ for a hard prior of $\Omega_\mathrm{m}=0.24$.

Our analysis differs from previous weak lensing analyses in three important aspects.  We correctly account for non-Gaussian sample variance using the method of \citet{Semboloni..0606648}, thus improving upon the purely Gaussian contribution given by \citet{2002A&A...396....1S}.  Using the results from STEP \citep{2007MNRAS.376...13M} we correct for the shear calibration bias and marginalise over our uncertainty in this correction. In addition we use the largest deep photometric redshift catalogue in existence \citep{Ilbert..0603217} to provide accurate models for the redshift distribution of sources; we also account for the effects of sample variance in these distributions \citep{Waerbeke..0603696}.  

Accounting for the non-Gaussian contribution to the shear covariance matrix, which dominates on small scales, is very important. The non-Gaussian contribution is about twice that of the Gaussian contribution alone at a scale of 10 arcminutes, this discrepancy increases to about an order of magnitude at 2 arcminutes. This increases the errors on the shear correlation function at small scales, leading to slightly weaker constraints on cosmology. Ideally we would estimate the shear covariance directly from the data, as is done in the previous analyses for both GaBoDS \citep{Hetterscheidt..0606571} and RCS \citep{2002ApJ...577..595H}. However, since we can not accomplish this for all the surveys in this work, due to a deficit of independent fields, we opt for a consistent approach by using the analytic treatment along with the non-Gaussian calibration as described in \citet{Semboloni..0606648}. 

\begin{figure}
\begin{center}
\includegraphics[scale=0.34]{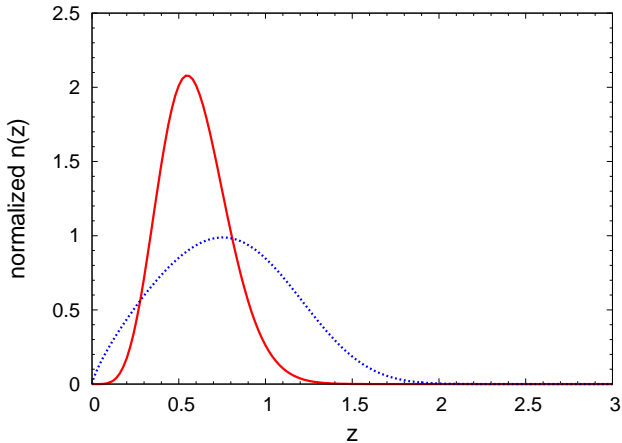}
\caption{\label{fig:rcs_nz} Redshift distributions for the RCS survey. The solid line shows the best fit $n(z)$ from \citet{2002ApJ...577..595H}, the dashed curve is our best fit $n(z)$, the average redshifts are 0.6 and 0.78 respectively.}
\end{center}
\end{figure}

Accurate determination of the redshift distribution of the sources is crucial for weak lensing cosmology since it is strongly degenerate with cosmological parameters. Past studies using small external photometric catalogues (such as the HDF) suffered from sample variance, a previously unquantified source of error that is taken into account here using the prescription of \citet{Waerbeke..0603696}. In most cases the revised redshift distributions were in reasonable agreement with previous results, however this was not the case for the RCS survey. We found that the previous estimation was biased toward low redshift (see Figure~\ref{fig:rcs_nz}), resulting in a significantly larger estimation of $\sigma_\mathrm{8}$ ($0.86^{+0.04}_{-0.05}$ for $\Omega_\mathrm{m}=0.3$) than is presented here ($0.69\pm0.07$). This difference is primarily due to a problem in the filter set conversion, between those used to image HDF (F300W, F450W, F606W, and F814W) and the Cousins $R_\mathrm{C}$ filter used by RCS. The less severe changes in $\sigma_{\rm 8}$ for the other surveys, shown in Figure~\ref{fig:compare0.24}, can also be largely understood from the updated redshift distributions. Comparing the median redshifts from the previous results (Table~\ref{tab:summary}) to those presented here (Table~\ref{redshiftfits}) there exists a clear trend associating increases in median redshift to decreases in estimated $\sigma_\mathrm{8}$ (see Figure~\ref{fig:compare0.24}) and vice-versa. Our estimate of median redshift for CFHTLS-Wide increases by $\sim10\pct$ and a corresponding decrease in $\sigma_\mathrm{8}$ is seen, a similar correspondence is seen for VIRMOS-DESCART where the decreased median redshift in this analysis results in a proportionately larger value of $\sigma_{\rm 8}$. The change in $\sigma_\mathrm{8}$ for GaBoDS, however, can not be understood by comparing median redshifts since the previous results use a piece wise function to fit the redshift distribution, hence comparing median redshifts is not meaningful.

Our revised cosmological constraints introduce tension, at the $\sim 2 \sigma$ level, between the results from the VIRMOS-DESCART and RCS survey as shown in Figure~\ref{fig:compare0.24}.  In this analysis we have accounted for systematic errors associated with shear measurement but have neglected potential systematics arising from correlations between galaxy shape and the underlying density field.  This is valid in the case of intrinsic galaxy alignments which are expected to contribute less than a percent of the cosmic shear signal for the deep surveys used in this analysis \citep{2006MNRAS.371..750H}.  What is currently uncertain however is the level of systematic error that arises from shear-ellipticity correlations \citep{HS04,2006MNRAS.371..750H,RM05,2007astro.ph..1671H} which could reduce the amplitude of the measured shear correlation function by $\sim 10\%$ \citep{2006MNRAS.371..750H}. It has been found from the analysis of the Sloan Digital Sky Survey \citep{RM05,2007astro.ph..1671H}, that different morphological galaxy types contribute differently to this effect.  It is therefore possible that the different R-band imaging of the RCS and the slightly lower median survey redshift make it more suceptible to this type of systematic error. Without complete redshift information for each survey, however, it is not possible to test this hypothesis or to correct for this potential source of error.  In the future deep multi-colour data will permit further investigation and correction for this potential source of systematic error.

In addition to using the best available photometric redshifts, we marginalise over the redshift distribution by selecting parameter triplets from their full 3D probability distribution, instead of fixing two parameters and varying the third as is often done. Thus the marginalisation is representative of the full range of $n(z)$ shapes, which are difficult to probe by varying one parameter due to degeneracies. 

We have conducted our analysis using two different functional forms for the redshift distribution, the standard form given in Eq.(\ref{brainerd}) and a new form given by Eq.(\ref{heymans}). The new function is motivated by the presence of a high-z tail when the entire photometric catalogue ($0.0 \leq z_\mathrm{p} \leq 4.0$) is used to estimate the redshift distribution, Eq.(\ref{brainerd}) does a poor job of fitting this distribution (Figure~\ref{widedist}). However when we restrict the photometric catalogue to the high confidence region ($0.2 \leq z_\mathrm{p} \leq 1.5$) both functions fit well, and the tendency for Eq.(\ref{heymans}) to exhibit a tail towards high-z increases the median redshift resulting in a slightly lower estimate of $\sigma_\mathrm{8}$. Though the tail of the distribution has only a small fraction of the total galaxies, they may have a significant lensing signal owing to their large redshifts. The influences of the different redshift distributions on cosmology is consistent within our $1\sigma$ errors but as survey sizes grow and statistical noise decreases, such differences will become significant.   As the CFHTLS-Deep will be the largest deep photometric redshift catalogue for some years to come, this posses a serious challenge to future surveys attempting to do precision cosmology. To assess the extent of any high redshift tail, future surveys should strive to include photometric bands in the near infra-red, allowing for accurate redshift estimations beyond $z=1.5$.

For $\Omega_\mathrm{m}=0.24$, the combined results using Eq.(\ref{brainerd}) are $0.80\pm0.05$ and $0.84\pm0.05$ for the unrestricted and high confidence photometric redshifts respectively, using Eq.(\ref{heymans}) we find $0.77\pm0.05$ and $0.82\pm0.06$ (these results use the non-linear power spectrum given by \citet{2003MNRAS.341.1311S}).
For completeness we provide results for both the \citet{2003MNRAS.341.1311S} and \citet{1996MNRAS.280L..19P} non-linear power spectra, the resulting $\sigma_\mathrm{8}$ values differ by $\sim3\pct$ of the \citet{2003MNRAS.341.1311S} value which is consistently the smaller of the two. The \citet{2003MNRAS.341.1311S} study is known to provide a more accurate estimation of the non-linear power than that of \citet{1996MNRAS.280L..19P}, for this reason we prefer the results obtained using the \citet{2003MNRAS.341.1311S} model. However, given the magnitude of variations resulting from different redshift distributions, this difference is not an important issue in the current work. Accurately determining the non-linear matter power spectrum is another challenge for future lensing surveys intent on precision cosmology. Alternatively, surveys focusing only on the large scale measurement of the shear (Fu et al., in prep) are able to avoid complications arising from the estimation of non-linear power on small scales. 

Surveys of varying depth provide different joint constraints in the $\Omega_\mathrm{m}-\sigma_\mathrm{8}$ plane, thus combining their likelihoods produces some degeneracy breaking. Taking the preferred prescription for the non-linear power spectrum, and using Eq.(\ref{brainerd}) along with the high confidence photometric redshifts to estimate the redshift distribution, we find an upper limit of $\Omega_\mathrm{m} <\sim 0.4$ and a lower limit of $\sigma_\mathrm{8} >\sim 0.6$ both at the $1\sigma$ level (see Figure~\ref{Fig:combinedconst}).

Weak lensing by large scale structure is an excellent means of constraining cosmology. The unambiguous interpretation of the shear signal allows for a direct measure of the dark matter power spectrum, allowing for a unique and powerful means of constraining cosmology. Our analysis has shown that accurately describing the redshift distribution of the sources is vital to future surveys intent on precision cosmology. Including near-IR bands to photometric redshift estimates will be a crucial step in achieving this goal, allowing for reliable redshift estimates at $z > 1.5$ and assessing the extent of a high-z tail. 

\section*{acknowledgments}
We would like to thank to the referee for very prompt and useful comments.
JB is supported by the Natural Sciences and Engineering Research Council (NSERC), and the Canadian Institute for Advanced Research (CIAR). CH acknowledges the support of the European Commission Programme 6$^\mathrm{th}$ frame work, Marie Curie Outgoing International Fellowship, contract number M01F-CT-2006-21891, and a CITA National Fellowship. ES thanks the hospitality of the University of British Columbia, which made this collaboration possible. LVW and HH are supported by NSERC, CIAR and the Canadian Foundation for Innovation (CFI).  YM thanks the Alexander Humboldt Foundation and the Terapix data center for support, and the AIfA for hospitality. This work is partly based on observations obtained with MegaPrime equipped with MegaCam, a joint project of CFHT and CEA/DAPNIA, at the Canada-France-Hawaii Telescope (CFHT) which is operated by the National Research Council (NRC) of Canada, the Institut National des Science de l'Univers of the Centre National de la Recherche Scientifique (CNRS) of France, and the University of Hawaii. This work is based in part on data products produced at
TERAPIX and the Canadian Astronomy Data Centre as part of the Canada-France-Hawaii Telescope Legacy Survey, a collaborative project of NRC and CNRS. This paper makes use of photometric redshifts produced jointly by Terapix and VVDS teams. This research was performed with infrastructure funded by the Canadian Foundation for Innovation and the British Columbia Knowledge Development Fund (A Parallel Computer for Compact-Object Physics). This research has been enabled by the use of WestGrid computing resources, which are funded in part by the Canada Foundation for Innovation, Alberta Innovation and Science, BC Advanced Education, the participating research institutions. WestGrid equipment is provided by IBM, Hewlett Packard and SGI.

\bibliographystyle{benjamin_bibsty}

\appendix
\section{Shear Correlation Function and Covariance Mtrix for the Surveys}
\label{appendixa}
We present both the measured shear correlation function and the covariance matrix for each survey. The shear correlation function, given in Tables \ref{tab:wideEB}, \ref{tab:gaboEB}, \ref{tab:rcsEB} and \ref{tab:virmEB} for the CFHTLS-Wide, GaBoDS, RCS and VIRMOS-DESCART surveys respectively, has been calibrated on large scales where $\langle M_{\rm ap}^2\left(\Delta\theta\right)\rangle_\perp$ is consistent with zero as described in \S4.

Tables \ref{tab:wideCOV}, \ref{tab:gaboCOV}, \ref{tab:rcsCOV} and \ref{tab:virmCOV} for the CFHTLS-Wide, GaBoDS, RCS and VIRMOS-DESCART surveys respectively tabulate the correlation coefficient matrix:
\begin{equation}
 \rg_{ij}=\frac{\Cg_{ij}}{\langle \xi_i^2 \rangle^{1/2}\langle \xi_j^2 \rangle^{1/2}},
\end{equation}
where $\Cg$ is the covariance matrix for each survey, as described in \S6. We also tabulate $\langle \xi_i^2 \rangle$ so that the covariance matrix may be calculated from the correlation coefficient matrix. Note that $\langle \xi_i^2 \rangle$ is the variance of the $i^{th}$ scale, equivalent to $\Cg_{ii}$.

\begin{table}
\caption{E and B modes of the shear correlation function for the CFHTLS-Wide survey, the error ($\delta\xi$) is statistical only, and given as the standard deviation.}
\label{tab:wideEB}
\begin{center}
\begin{tabular}{cccc}
\hline
\hline
$\theta$ (arcmin) & $\xi_E$ & $\xi_B$ & $\delta\xi$\\
\hline
0.74 & 1.53060e-04 & -4.16557e-05 & 2.31000e-05 \\
1.44 & 1.00161e-04 & 1.79443e-05 & 1.22000e-05 \\
2.37 & 5.82357e-05 & -1.60557e-05 & 9.64000e-06 \\
3.54 & 3.11136e-05 & 7.14429e-06 & 6.66000e-06 \\
5.41 & 2.89396e-05 & -2.05571e-06 & 4.29000e-06 \\
7.28 & 3.08030e-05 & -3.47571e-06 & 4.76000e-06 \\
8.69 & 1.41053e-05 & 6.14429e-06 & 4.43000e-06 \\
11.72 & 1.66208e-05 & -5.15714e-07 & 2.16000e-06 \\
18.74 & 1.50680e-05 & 4.14286e-07 & 1.25000e-06 \\
28.09 & 1.07719e-05 & -7.95715e-07 & 1.03000e-06 \\
37.44 & 8.72227e-06 & 4.14286e-07 & 9.10000e-07 \\
49.12 & 8.11150e-06 & -2.18571e-06 & 6.64000e-07 \\
62.92 & 8.97071e-06 & -3.01571e-06 & 6.25000e-07 \\
\hline
\hline
\end{tabular}
\end{center}
\end{table}

\begin{table}
\caption{E and B modes of the shear correlation function for GaBoDS, the error ($\delta\xi$) is statistical only, and given as the standard deviation.}
\label{tab:gaboEB}
\begin{center}
\begin{tabular}{cccc}
\hline
\hline
$\theta$ (arcmin) & $\xi_E$ & $\xi_B$ & $\delta\xi$\\
\hline
0.36 & 2.80637e-04 & 1.38754e-04 & 9.54000e-05 \\
0.50 & 1.72930e-04 & 1.24754e-04 & 6.92000e-05 \\
0.70 & 1.15363e-04 & 6.74540e-05 & 4.99000e-05 \\
0.98 & 1.20006e-04 & -1.59760e-05 & 3.81000e-05 \\
1.36 & 8.75079e-05 & 2.76540e-05 & 2.80000e-05 \\
1.90 & 7.43231e-05 & 8.35400e-06 & 2.15000e-05 \\
2.65 & 9.55859e-05 & -2.30860e-05 & 1.71000e-05 \\
3.69 & 6.48524e-05 & 3.05400e-06 & 1.40000e-05 \\
5.16 & 4.30325e-05 & 5.25400e-06 & 1.19000e-05 \\
7.19 & 2.73408e-05 & -7.24600e-06 & 1.03000e-05 \\
10.03 & 9.97772e-06 & -7.46000e-07 & 8.65000e-06 \\
13.99 & 1.22154e-05 & -6.54600e-06 & 6.55000e-06 \\
\hline
\hline
\end{tabular}
\end{center}
\end{table}

\begin{table}
\caption{E and B modes of the shear correlation function for RCS, the error ($\delta\xi$) is statistical only, and given as the standard deviation.}
\label{tab:rcsEB}
\begin{center}
\begin{tabular}{cccc}
\hline
\hline
$\theta$ (arcmin) & $\xi_E$ & $\xi_B$ & $\delta\xi$\\
\hline
1.12 & 9.71718e-05 & 2.78320e-05 & 1.99000e-05 \\
1.75 & 3.38178e-05 & 1.67320e-05 & 1.91000e-05 \\
2.50 & 3.74410e-05 & 4.00200e-06 & 1.16000e-05 \\
3.24 & 4.27205e-05 & 6.63200e-06 & 1.43000e-05 \\
4.00 & 1.49772e-05 & 1.58320e-05 & 9.21000e-06 \\
5.75 & 1.52878e-05 & -8.80000e-08 & 5.13000e-06 \\
8.50 & 1.50807e-05 & -3.89600e-06 & 3.95000e-06 \\
12.25 & 6.89234e-06 & 2.15200e-06 & 2.85000e-06 \\
16.49 & 4.81573e-06 & 1.83200e-06 & 2.65000e-06 \\
20.74 & 4.69151e-06 & 1.82200e-06 & 2.30000e-06 \\
26.49 & 5.29607e-06 & 3.59200e-06 & 1.75000e-06 \\
34.99 & 5.33747e-06 & 1.11200e-06 & 1.36000e-06 \\
46.49 & 6.91304e-06 & -4.88000e-07 & 1.08000e-06 \\
56.51 & 6.69565e-06 & -4.88000e-07 & 1.34000e-06 \\
\hline
\hline
\end{tabular}
\end{center}
\end{table}

\begin{table}
\caption{E and B modes of the shear correlation function for the VIRMOS-DESCART survey, the error ($\delta\xi$) is statistical only, and given as the standard deviation.}
\label{tab:virmEB}
\begin{center}
\begin{tabular}{cccc}
\hline
\hline
$\theta$ (arcmin) & $\xi_E$ & $\xi_B$ & $\delta\xi$\\
\hline
0.73 & 2.26695e-04 & 3.91133e-05 & 3.98000e-05 \\
1.05 & 1.69759e-04 & -1.59667e-06 & 3.90000e-05 \\
1.51 & 1.30421e-04 & 5.31333e-06 & 2.20000e-05 \\
2.29 & 9.60525e-05 & 2.82133e-05 & 1.57000e-05 \\
3.20 & 6.60318e-05 & -1.38767e-05 & 1.35000e-05 \\
4.11 & 5.56798e-05 & 6.41333e-06 & 1.22000e-05 \\
5.02 & 5.29883e-05 & 3.71333e-06 & 1.12000e-05 \\
6.39 & 3.50794e-05 & -1.26667e-06 & 7.46000e-06 \\
8.21 & 1.94893e-05 & 2.11333e-06 & 6.75000e-06 \\
11.41 & 2.72119e-05 & -8.46667e-07 & 3.99000e-06 \\
15.97 & 1.98620e-05 & 4.81333e-06 & 3.50000e-06 \\
20.53 & 1.67357e-05 & 5.41333e-06 & 3.23000e-06 \\
25.25 & 1.05659e-05 & 4.51333e-06 & 2.95000e-06 \\
30.23 & 9.77399e-06 & 1.31333e-06 & 2.73000e-06 \\
35.10 & 1.06073e-05 & -1.05667e-06 & 2.71000e-06 \\
39.78 & 1.17771e-05 & -2.08667e-06 & 2.61000e-06 \\
\hline
\hline
\end{tabular}
\end{center}
\end{table}

\begin{table*}
\caption{The correlation coefficient matrix ($\rg$) for the CFHTLS-Wide survey, the scales $\theta_i$ correspond to those given in Table~\ref{tab:wideEB}. The column $\langle \xi_i^2 \rangle$ is given in units of $10^{-10}$, and can be used to reconstruct the covariance matrix $\Cg$.}
\label{tab:wideCOV}
\begin{center}
\begin{tabular}{ccccccccccccccc}
\hline
\hline
& $\langle \xi_i^2 \rangle$ & $\theta_1$ & $\theta_2$ & $\theta_3$ & $\theta_4$ & $\theta_5$ & $\theta_6$ & $\theta_7$ & $\theta_8$ & $\theta_9$ & $\theta_{10}$ & $\theta_{11}$ & $\theta_{12}$ & $\theta_{13}$\\
\hline
$\theta_{1}$ & 29.9246 &  1.0000 &  0.2595 &  0.1790 &  0.1975 &  0.1900 &  0.1223 &  0.0841 &  0.1025 &  0.0707 &  0.0587 &  0.0549 &  0.0393 &  0.0266 \\ 
$\theta_{2}$ &  6.5960 &  0.2595 &  1.0000 &  0.2175 &  0.2387 &  0.2330 &  0.1519 &  0.1062 &  0.1527 &  0.1431 &  0.1251 &  0.1169 &  0.0837 &  0.0567 \\ 
$\theta_{3}$ &  4.2569 &  0.1790 &  0.2175 &  1.0000 &  0.2182 &  0.2192 &  0.1585 &  0.1243 &  0.1904 &  0.1784 &  0.1560 &  0.1457 &  0.1044 &  0.0707 \\ 
$\theta_{4}$ &  1.3550 &  0.1975 &  0.2387 &  0.2182 &  1.0000 &  0.3734 &  0.2861 &  0.2228 &  0.3405 &  0.3180 &  0.2778 &  0.2594 &  0.1855 &  0.1254 \\ 
$\theta_{5}$ &  0.5285 &  0.1900 &  0.2330 &  0.2192 &  0.3734 &  1.0000 &  0.4829 &  0.3683 &  0.5554 &  0.5091 &  0.4370 &  0.4120 &  0.2970 &  0.2006 \\ 
$\theta_{6}$ &  0.6121 &  0.1223 &  0.1519 &  0.1585 &  0.2861 &  0.4829 &  1.0000 &  0.3505 &  0.5277 &  0.4732 &  0.3984 &  0.3790 &  0.2757 &  0.1865 \\ 
$\theta_{7}$ &  0.8166 &  0.0841 &  0.1062 &  0.1243 &  0.2228 &  0.3683 &  0.3505 &  1.0000 &  0.4683 &  0.4119 &  0.3464 &  0.3293 &  0.2393 &  0.1615 \\ 
$\theta_{8}$ &  0.2533 &  0.1025 &  0.1527 &  0.1904 &  0.3405 &  0.5554 &  0.5277 &  0.4683 &  1.0000 &  0.7518 &  0.6322 &  0.5951 &  0.4289 &  0.2901 \\ 
$\theta_{9}$ &  0.1796 &  0.0707 &  0.1431 &  0.1784 &  0.3180 &  0.5091 &  0.4732 &  0.4119 &  0.7518 &  1.0000 &  0.7832 &  0.7144 &  0.5112 &  0.3522 \\ 
$\theta_{10}$ &  0.1447 &  0.0587 &  0.1251 &  0.1560 &  0.2778 &  0.4370 &  0.3984 &  0.3464 &  0.6322 &  0.7832 &  1.0000 &  0.8248 &  0.5722 &  0.3883 \\ 
$\theta_{11}$ &  0.1074 &  0.0549 &  0.1169 &  0.1457 &  0.2594 &  0.4120 &  0.3790 &  0.3293 &  0.5951 &  0.7144 &  0.8248 &  1.0000 &  0.6889 &  0.4453 \\ 
$\theta_{12}$ &  0.1274 &  0.0393 &  0.0837 &  0.1044 &  0.1855 &  0.2970 &  0.2757 &  0.2393 &  0.4289 &  0.5112 &  0.5722 &  0.6889 &  1.0000 &  0.4240 \\ 
$\theta_{13}$ &  0.1615 &  0.0266 &  0.0567 &  0.0707 &  0.1254 &  0.2006 &  0.1865 &  0.1615 &  0.2901 &  0.3522 &  0.3883 &  0.4453 &  0.4240 &  1.0000 \\ 
\hline
\hline
\end{tabular}
\end{center}
\end{table*}

\begin{table*}
\caption{The correlation coefficient matrix ($\rg$) for GaBoDS, the scales $\theta_i$ correspond to those given in Table~\ref{tab:gaboEB}. The column $\langle \xi_i^2 \rangle$ is given in units of $10^{-10}$, and can be used to reconstruct the covariance matrix $\Cg$.}
\label{tab:gaboCOV}
\begin{center}
\begin{tabular}{cccccccccccccc}
\hline
\hline
& $\langle \xi_i^2 \rangle$ & $\theta_1$ & $\theta_2$ & $\theta_3$ & $\theta_4$ & $\theta_5$ & $\theta_6$ & $\theta_7$ & $\theta_8$ & $\theta_9$ & $\theta_{10}$ & $\theta_{11}$ & $\theta_{12}$\\
\hline
$\theta_{1}$ & 323.7553 &  1.0000 &  0.1199 &  0.1298 &  0.1737 &  0.1383 &  0.1627 &  0.1028 &  0.1404 &  0.1148 &  0.0890 &  0.0868 &  0.0590 \\ 
$\theta_{2}$ & 223.3641 &  0.1199 &  1.0000 &  0.1214 &  0.1566 &  0.1249 &  0.1490 &  0.0958 &  0.1336 &  0.1117 &  0.0892 &  0.0896 &  0.0604 \\ 
$\theta_{3}$ & 80.2861 &  0.1298 &  0.1214 &  1.0000 &  0.2051 &  0.1608 &  0.1937 &  0.1269 &  0.1810 &  0.1552 &  0.1284 &  0.1351 &  0.0881 \\ 
$\theta_{4}$ & 22.1309 &  0.1737 &  0.1566 &  0.2051 &  1.0000 &  0.2487 &  0.2988 &  0.1991 &  0.2902 &  0.2548 &  0.2150 &  0.2286 &  0.1514 \\ 
$\theta_{5}$ & 18.2583 &  0.1383 &  0.1249 &  0.1608 &  0.2487 &  1.0000 &  0.2817 &  0.1891 &  0.2804 &  0.2536 &  0.2193 &  0.2327 &  0.1654 \\ 
$\theta_{6}$ &  6.9091 &  0.1627 &  0.1490 &  0.1937 &  0.2988 &  0.2817 &  1.0000 &  0.2815 &  0.4205 &  0.3819 &  0.3349 &  0.3720 &  0.2681 \\ 
$\theta_{7}$ &  9.2132 &  0.1028 &  0.0958 &  0.1269 &  0.1991 &  0.1891 &  0.2815 &  1.0000 &  0.3416 &  0.3155 &  0.2839 &  0.3190 &  0.2314 \\ 
$\theta_{8}$ &  2.6680 &  0.1404 &  0.1336 &  0.1810 &  0.2902 &  0.2804 &  0.4205 &  0.3416 &  1.0000 &  0.5807 &  0.5259 &  0.5879 &  0.4247 \\ 
$\theta_{9}$ &  2.1958 &  0.1148 &  0.1117 &  0.1552 &  0.2548 &  0.2536 &  0.3819 &  0.3155 &  0.5807 &  1.0000 &  0.5801 &  0.6401 &  0.4606 \\ 
$\theta_{10}$ &  2.0117 &  0.0890 &  0.0892 &  0.1284 &  0.2150 &  0.2193 &  0.3349 &  0.2839 &  0.5259 &  0.5801 &  1.0000 &  0.6561 &  0.4734 \\ 
$\theta_{11}$ &  1.1340 &  0.0868 &  0.0896 &  0.1351 &  0.2286 &  0.2327 &  0.3720 &  0.3190 &  0.5879 &  0.6401 &  0.6561 &  1.0000 &  0.6095 \\ 
$\theta_{12}$ &  1.1960 &  0.0590 &  0.0604 &  0.0881 &  0.1514 &  0.1654 &  0.2681 &  0.2314 &  0.4247 &  0.4606 &  0.4734 &  0.6095 &  1.0000 \\ 
\hline
\hline
\end{tabular}
\end{center}
\end{table*}

\begin{table*}
\caption{The correlation coefficient matrix ($\rg$) for RCS, the scales $\theta_i$ correspond to those given in Table~\ref{tab:rcsEB}. The column $\langle \xi_i^2 \rangle$ is given in units of $10^{-10}$, and can be used to reconstruct the covariance matrix $\Cg$.}
\label{tab:rcsCOV}
\begin{center}
\begin{tabular}{cccccccccccccccc}
\hline
\hline
& $\langle \xi_i^2 \rangle$ & $\theta_1$ & $\theta_2$ & $\theta_3$ & $\theta_4$ & $\theta_5$ & $\theta_6$ & $\theta_7$ & $\theta_8$ & $\theta_9$ & $\theta_{10}$ & $\theta_{11}$ & $\theta_{12}$ & $\theta_{13}$ & $\theta_{14}$\\
\hline
$\theta_{1}$ & 12.9511 &  1.0000 &  0.0957 &  0.1251 &  0.0775 &  0.0543 &  0.1070 &  0.0731 &  0.0695 &  0.0618 &  0.0578 &  0.0416 &  0.0535 &  0.0484 &  0.0389 \\ 
$\theta_{2}$ &  7.0289 &  0.0957 &  1.0000 &  0.1354 &  0.0822 &  0.0574 &  0.1155 &  0.0832 &  0.0885 &  0.0832 &  0.0784 &  0.0564 &  0.0725 &  0.0656 &  0.0526 \\ 
$\theta_{3}$ &  1.8431 &  0.1251 &  0.1354 &  1.0000 &  0.1445 &  0.0996 &  0.2076 &  0.1575 &  0.1721 &  0.1626 &  0.1529 &  0.1100 &  0.1414 &  0.1278 &  0.1024 \\ 
$\theta_{4}$ &  2.7380 &  0.0775 &  0.0822 &  0.1445 &  1.0000 &  0.0788 &  0.1696 &  0.1290 &  0.1407 &  0.1333 &  0.1254 &  0.0902 &  0.1159 &  0.1048 &  0.0840 \\ 
$\theta_{5}$ &  3.5622 &  0.0543 &  0.0574 &  0.0996 &  0.0788 &  1.0000 &  0.1535 &  0.1149 &  0.1229 &  0.1161 &  0.1099 &  0.0791 &  0.1016 &  0.0919 &  0.0736 \\ 
$\theta_{6}$ &  0.4239 &  0.1070 &  0.1155 &  0.2076 &  0.1696 &  0.1535 &  1.0000 &  0.3459 &  0.3614 &  0.3391 &  0.3192 &  0.2293 &  0.2949 &  0.2667 &  0.2136 \\ 
$\theta_{7}$ &  0.4406 &  0.0731 &  0.0832 &  0.1575 &  0.1290 &  0.1149 &  0.3459 &  1.0000 &  0.3797 &  0.3443 &  0.3159 &  0.2261 &  0.2904 &  0.2631 &  0.2108 \\ 
$\theta_{8}$ &  0.2373 &  0.0695 &  0.0885 &  0.1721 &  0.1407 &  0.1229 &  0.3614 &  0.3797 &  1.0000 &  0.4902 &  0.4191 &  0.3076 &  0.4010 &  0.3645 &  0.2920 \\ 
$\theta_{9}$ &  0.1925 &  0.0618 &  0.0832 &  0.1626 &  0.1333 &  0.1161 &  0.3391 &  0.3443 &  0.4902 &  1.0000 &  0.4900 &  0.3514 &  0.4521 &  0.4105 &  0.3305 \\ 
$\theta_{10}$ &  0.1738 &  0.0578 &  0.0784 &  0.1529 &  0.1254 &  0.1099 &  0.3192 &  0.3159 &  0.4191 &  0.4900 &  1.0000 &  0.3847 &  0.4827 &  0.4361 &  0.3529 \\ 
$\theta_{11}$ &  0.2279 &  0.0416 &  0.0564 &  0.1100 &  0.0902 &  0.0791 &  0.2293 &  0.2261 &  0.3076 &  0.3514 &  0.3847 &  1.0000 &  0.4451 &  0.3832 &  0.3043 \\ 
$\theta_{12}$ &  0.0848 &  0.0535 &  0.0725 &  0.1414 &  0.1159 &  0.1016 &  0.2949 &  0.2904 &  0.4010 &  0.4521 &  0.4827 &  0.4451 &  1.0000 &  0.6601 &  0.5100 \\ 
$\theta_{13}$ &  0.0578 &  0.0484 &  0.0656 &  0.1278 &  0.1048 &  0.0919 &  0.2667 &  0.2631 &  0.3645 &  0.4105 &  0.4361 &  0.3832 &  0.6601 &  1.0000 &  0.6379 \\ 
$\theta_{14}$ &  0.0540 &  0.0389 &  0.0526 &  0.1024 &  0.0840 &  0.0736 &  0.2136 &  0.2108 &  0.2920 &  0.3305 &  0.3529 &  0.3043 &  0.5100 &  0.6379 &  1.0000 \\
\hline
\hline
\end{tabular}
\end{center}
\end{table*}

\begin{table*}
\caption{The correlation coefficient matrix ($\rg$) for the VIRMOS-DESCART survey, the scales $\theta_i$ correspond to those given in Table~\ref{tab:virmEB}. The column $\langle \xi_i^2 \rangle$ is given in units of $10^{-10}$, and can be used to reconstruct the covariance matrix $\Cg$.}
\label{tab:virmCOV}
\begin{center}
\begin{tabular}{cccccccccc}
\hline
\hline
& $\langle \xi_i^2 \rangle$ & $\theta_1$ & $\theta_2$ & $\theta_3$ & $\theta_4$ & $\theta_5$ & $\theta_6$ & $\theta_7$ & $\theta_8$\\
\hline
$\theta_{1}$ & 46.5520 &  1.0000 &  0.3477 &  0.3669 &  0.1949 &  0.2149 &  0.2176 &  0.1959 &  0.1922 \\ 
$\theta_{2}$ & 23.0214 &  0.3477 &  1.0000 &  0.4015 &  0.2029 &  0.2248 &  0.2352 &  0.2152 &  0.2135 \\
$\theta_{3}$ &  8.9212 &  0.3669 &  0.4015 &  1.0000 &  0.2598 &  0.2839 &  0.2849 &  0.2575 &  0.2565 \\
$\theta_{4}$ & 12.1714 &  0.1949 &  0.2029 &  0.2598 &  1.0000 &  0.1862 &  0.1855 &  0.1706 &  0.1770 \\
$\theta_{5}$ &  4.7621 &  0.2149 &  0.2248 &  0.2839 &  0.1862 &  1.0000 &  0.2584 &  0.2498 &  0.2838 \\
$\theta_{6}$ &  2.6912 &  0.2176 &  0.2352 &  0.2849 &  0.1855 &  0.2584 &  1.0000 &  0.3461 &  0.3858 \\
$\theta_{7}$ &  2.1129 &  0.1959 &  0.2152 &  0.2575 &  0.1706 &  0.2498 &  0.3461 &  1.0000 &  0.4465 \\
$\theta_{8}$ &  1.2564 &  0.1922 &  0.2135 &  0.2565 &  0.1770 &  0.2838 &  0.3858 &  0.4465 &  1.0000 \\
$\theta_{9}$ &  1.0789 &  0.1629 &  0.1738 &  0.2156 &  0.1665 &  0.2690 &  0.3636 &  0.4161 &  0.5497 \\
$\theta_{10}$ &  0.6759 &  0.1383 &  0.1492 &  0.2110 &  0.1780 &  0.2857 &  0.3791 &  0.4290 &  0.5609 \\
$\theta_{11}$ &  0.7749 &  0.0843 &  0.1029 &  0.1652 &  0.1415 &  0.2259 &  0.2962 &  0.3349 &  0.4407 \\
$\theta_{12}$ &  0.8042 &  0.0648 &  0.0921 &  0.1477 &  0.1263 &  0.2017 &  0.2681 &  0.3026 &  0.3929 \\
$\theta_{13}$ &  0.6214 &  0.0629 &  0.0893 &  0.1432 &  0.1224 &  0.1954 &  0.2597 &  0.2932 &  0.3804 \\
$\theta_{14}$ &  0.4127 &  0.0672 &  0.0954 &  0.1529 &  0.1306 &  0.2084 &  0.2771 &  0.3127 &  0.4056 \\
$\theta_{15}$ &  0.3427 &  0.0647 &  0.0917 &  0.1469 &  0.1255 &  0.2003 &  0.2664 &  0.3007 &  0.3903 \\
$\theta_{16}$ &  0.3482 &  0.0574 &  0.0813 &  0.1303 &  0.1112 &  0.1776 &  0.2362 &  0.2668 &  0.3464 \\
\hline
\hline
& $\theta_9$ & $\theta_{10}$ & $\theta_{11}$ & $\theta_{12}$ & $\theta_{13}$ & $\theta_{14}$ & $\theta_{15}$ & $\theta_{16}$ & \\
\hline
$\theta_{1}$ & 0.1629 & 0.1383 & 0.0843 & 0.0648 & 0.0629 & 0.0672 & 0.0647 & 0.0574 & \\
$\theta_{2}$ & 0.1738 & 0.1492 & 0.1029 & 0.0921 & 0.0893 & 0.0954 & 0.0917 & 0.0813 & \\
$\theta_{3}$ & 0.2156 & 0.2110 & 0.1652 & 0.1477 & 0.1432 & 0.1529 & 0.1469 & 0.1303 & \\
$\theta_{4}$ & 0.1665 & 0.1780 & 0.1415 & 0.1263 & 0.1224 & 0.1306 & 0.1255 & 0.1112 & \\
$\theta_{5}$ & 0.2690 & 0.2857 & 0.2259 & 0.2017 & 0.1954 & 0.2084 & 0.2003 & 0.1776 & \\
$\theta_{6}$ & 0.3636 & 0.3791 & 0.2962 & 0.2681 & 0.2597 & 0.2771 & 0.2664 & 0.2362 & \\
$\theta_{7}$ & 0.4161 & 0.4290 & 0.3349 & 0.3026 & 0.2932 & 0.3127 & 0.3007 & 0.2668 & \\
$\theta_{8}$ & 0.5497 & 0.5609 & 0.4407 & 0.3929 & 0.3804 & 0.4056 & 0.3903 & 0.3464 & \\
$\theta_{9}$ & 1.0000 & 0.6363 & 0.4908 & 0.4272 & 0.4124 & 0.4388 & 0.4227 & 0.3756 & \\
$\theta_{10}$ & 0.6363 & 1.0000 & 0.6363 & 0.5224 & 0.5159 & 0.5596 & 0.5404 & 0.4812 & \\
$\theta_{11}$ & 0.4908 & 0.6363 & 1.0000 & 0.4884 & 0.4864 & 0.5316 & 0.5136 & 0.4577 & \\
$\theta_{12}$ & 0.4272 & 0.5224 & 0.4884 & 1.0000 & 0.5001 & 0.5302 & 0.5107 & 0.4541 & \\
$\theta_{13}$ & 0.4124 & 0.5159 & 0.4864 & 0.5001 & 1.0000 & 0.6254 & 0.5957 & 0.5249 & \\
$\theta_{14}$ & 0.4388 & 0.5596 & 0.5316 & 0.5302 & 0.6254 & 1.0000 & 0.7429 & 0.6519 & \\
$\theta_{15}$ & 0.4227 & 0.5404 & 0.5136 & 0.5107 & 0.5957 & 0.7429 & 1.0000 & 0.6983 & \\
$\theta_{16}$ & 0.3756 & 0.4812 & 0.4577 & 0.4541 & 0.5249 & 0.6519 & 0.6983 & 1.0000 & \\
\hline
\hline
\end{tabular}
\end{center}
\end{table*}

\end{document}